\documentclass{easychair}
\usepackage{long2}
\usepackage{pdflscape}
\usepackage{longtable}
\usepackage{cite}
\usepackage{epstopdf}
\usepackage{dingbat}
\usepackage{pifont}
\newcommand{\cmark}{\ding{51}}%
\newcommand{\xmark}{\ding{55}}%

\begin{document}

%% Front Matter
%%
% Regular title as in the article class.
%
\title{Security of 5G-Mobile Backhaul Networks: A Survey}

% \titlerunning{} has to be set to either the main title or its shorter
% version for the running heads. Use {\sf} for highlighting your system
% name, application, or a tool.
%
\titlerunning{Security of 5G-Mobile Backhaul Networks: A Survey}

% For only the editors. Authors, please keep this commented out

% Authors are joined by \and and their affiliations are on the
% subsequent lines separated by \\ just like the article class
% allows.
%
\author{\\
Gaurav Choudhary\thanks{Corresponding author: Department of Information Security Engineering, Soonchunhyang University, Asan-si-31538, South Korea.
},~ Jiyoon Kim, ~Vishal Sharma\\
${}^{}$Department of Information Security Engineering, Soonchunhyang University, Asan-si-31538, South Korea\\
gauravchoudhary7777@gmail.com, 74jykim@gmail.com, vishal\_sharma2012@hotmail.com
}

% \authorrunning{} has to be set for the shorter version of the authors' names;
% otherwise a warning will be rendered in the running heads.
%
\authorrunning{Choudhary et al.}

\maketitle

%------------------------------------------------------------------------------
% Abstract
%
\begin{abstract}
The rapid involution of the mobile generation with incipient data networking capabilities and utilization has exponentially increased the data traffic volumes. Such traffic drains various key issues in 5G mobile backhaul networks. Security of mobile backhaul is of utmost importance; however, there are a limited number of articles, which have explored such a requirement. This paper discusses the potential design issues and key challenges of the secure 5G mobile backhaul architecture. The comparisons of the existing state-of-the-art solutions for secure mobile backhaul, together with their major contributions have been explored. Furthermore, the paper discussed various key issues related to Quality of Service (QoS), routing and scheduling, resource management, capacity enhancement, latency, security-management, and handovers using mechanisms like Software Defined Networking and millimeter Wave technologies. Moreover, the trails of research challenges and future directions are additionally presented. 
\end{abstract}
%\tableofcontents
%------------------------------------------------------------------------------
\section{Introduction}
The tremendous change in technologies and excessive demands of users open quality requirements related to data traffics and network load. Such demands are manageable through the new generation of mobile networks.  Following this, the 5G has captured the market while presenting itself as a significant candidate for new communication networks. 5G aims at providing high data rates, large capacity, better Quality of Service (QoS) at low latency and high reliability. Furthermore, it is becoming an active candidate for various applications such as public safety communication~\cite{naqvi2018drone}, massive IoT~\cite{li20185g}, smart home~\cite{sharma2018shsec}, smart city~\cite{dalla2018using}, smart health~\cite{dighriri2018big}, and smart grid~\cite{leligou2018smart}. The mobile operator facilitates various factors for research in the advancement of 5G technologies, which are actively supported by various standardizations. The active research industries and organizations, such as 5GPP, 3GPP, IETF, NGMN, Huawei, Samsung, Nokia, NEC, continuously focused on the performance of 5G for providing a more effective communication and connectivity to the end users~\cite{kumar20195g}.

\begin{figure}[ht!]
\centering
\includegraphics[width=350px]{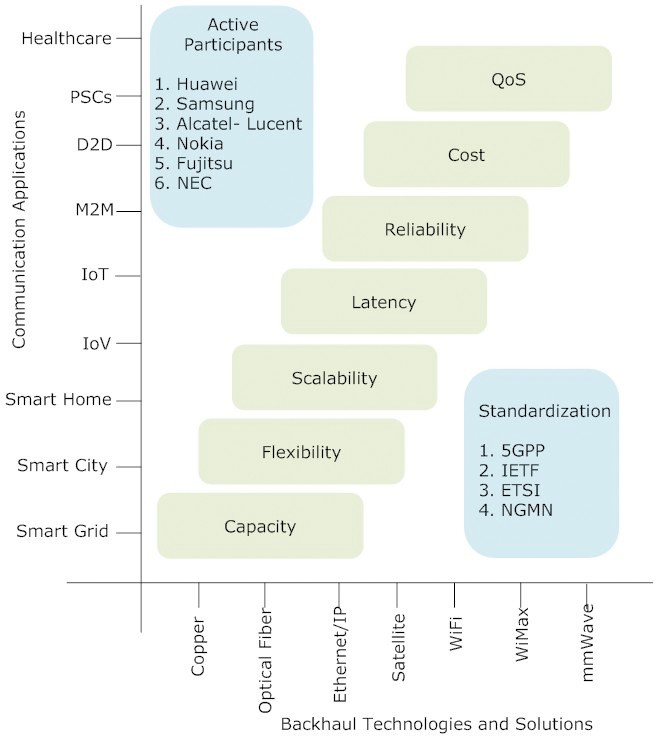}
\caption{The communication applications and backhaul solutions (wired and wireless) along with the network requirements.}
\label{technology}
\end{figure}

The mobile networks and operators have a basic necessity for a trusted network which incorporates the confidentiality, integrity, and availability of services in communications. It has been surveyed that the network component policies and architectural design play an important role in network reliability and sustainability~\cite{parvez2018survey}.  Furthermore, the traditional cellular networks are adopted widely by the network operators. However, due to the increasing traffic load, the network or mobile operator was forced to explore new generation technologies. As shown in Fig~\ref{technology}, the network requirements should maintain a tradeoff between applications and available solutions.

In addition to other features, the 5G mobile network demands a fast handover and low latency during mobile communications~\cite{costa2018sdn}. Backhaul plays a crucial role in providing such a facility. It is a bridge between the Radio Access Network (RAN) equipment and mobile backbone network. Backhaul is responsible for data transmission between the mobile modes and the network. The effective mobile backhaul increases bandwidth utilization and network efficiency~\cite{ustunbas2018adaptive}. The 5G mobile architecture incorporates radio networks and smart antennas for better performance. The next generation mobile communications also include heterogeneous networks and centralized architecture (C-RAN). The physical and mac layer advancements also improved the 5G communication for better performance. Majority of these advancements focused on the adaptive beamforming, PtmP, full duplex radio technologies, and directional mac protocols. Irrespective of these enhancements, the security issues are proportionally increasing due to various threats and vulnerabilities.

The mobile operator wants to maintain a tradeoff between the security solutions and implementation cost (Capital Expenditure(CapEx) and Operating Expenses(OpEx))~\cite{gomes2018cyber}~\cite{uchida2017mobile}. The network demands appropriate facilities and should employ new techniques such as Software-Defined Networking (SDN), Millimeter Wave (mmWave), or Satellite-based solutions to address backhaul issues. The SDN supports an effective and reliable traffic monitoring and management. The mmWave based implementation opens a new platform for mobile backhauls to provide a consistent support for a fast access to services. The massive MIMO are adopted in small cell scenarios and provides cooperation by deploying an antenna array that has active elements in excess of the number of users~\cite{jisis18-8-2-01}. In addition, it provides tremendous diversity gain and new aspects for network design to improve performance~\cite{alsharif2017evolution}.

Additionally, high performance and QoS services have to be emphasized on backhaul networks. Aside from these, some certain issues such as interference management, link utilization, reliability, flatness, and flexibility also need considerable attention. It is desired that an adequate solution should satisfy the exponential rise in user and traffic capacity of mobile broadband communications and should support flexible networks for a high data rate and large coverage~\cite{agiwal2016next}~\cite{siddiquee2018estimation}.

\begin{table}[!ht]
\caption{Abbreviation and Key terms.}
\begin{tabular}{llll}

\hline
\hline
\textbf{Abbreviation} & \textbf{Full Form}                & \textbf{Abbreviation} & \textbf{Full Form}               \\
 \hline
CapEx                 & Capital Expenditure               & NGMN                  & Next Generation Mobile Networks  \\
C-RAN                 & Centralized Radio Access Network  & NFV                   & Network Functions Virtualization \\
FSO                   & Free-Space Optical                & OpEx                  & Operating Expenses               \\
HetNets               & Heterogeneous Networks            & PTP                   & Point-to-Point                   \\
IETF                  & Internet Engineering Task Force   & PtmP                  & Point-to-Multipoint              \\
IP                    & Internet Protocol                 & QoS                   & Quality of Service               \\
MAC                   & Medium Access Control             & RAN                   & Radio Access Network             \\
mmWave                & Millimeter Wave                   & SDN                   & Software-Defined Networking      \\
MIDO                  & Multiple Input Distributed Output & TDM                   & Time-Division Multiplexing       \\
MIMO                  & Multiple Input Multiple Output & UE                    & User Equipment                   \\
LTE                   & Long-Term Evolution               & VPN                   & Virtual Private Network          \\
LOS                   & Line of Sight                     & WiBACK                & wireless Backhaul        \\
\hline
\end{tabular}
\end{table}

\subsection{Our Contributions}
The existing surveys emphasized the performance and architectural issues of 5G mobile backhauls. Majority of these only focused on the performance and architectural issues but left out the security aspects of 5G mobile backhaul. Therefore, the prime objective of this paper is to fill this gap by considering various key issues including security concerns and requirements related to 5G mobile backhaul. The major contributions of the survey are:
\begin{itemize}
\item The discussion of feasibility and challenges of existing 5G mobile architecture have been presented. The state-of-the-art comparison and road-map of existing surveys have been included while emphasizing on their major contributions.
\item The taxonomy of the 5G mobile backhaul framework and models including, general solutions, SDN, and the mmWave-based mechanisms have been discussed.
\item The backhaul security threats, attacks, vulnerability issues, and requirements for 5G mobile communication have been discussed.
\item The trails of major ongoing research challenges and future directions in the era of 5G mobile backhaul have been presented.
\end{itemize}

The rest of the article is structured as follows: Section 2 presents the existing surveys and roadmap of mobile backhauls. Section 3 discusses the 5G mobile backhaul and existing frameworks and solutions. Section 4 presents the backhaul security requirements and existing 5G mobile backhaul security solutions. Section 5 includes the research challenges and future directions. Finally, Section 6 concludes the article.

\section{Existing Surveys and Roadmap for Mobile Backhaul}
\label{sect:surveys}
The exponential increase in the number of mobile users and deployment of 5G technologies grab a huge attention towards the mobile backhaul. Focusing on mobile backhaul, numerous research surveys have been published to emphasize the challenges, design issues, and various future research with the perspective of architecture, deployment, and performance considerations as discussed in the Table~\ref{Table1}. Saha et al.~\cite{saha2016evolution} presented a survey on the technology evolution of 5G mobile networks while considering network backhaul and synchronization issues. The authors emphasized the 5G network architecture while considering radio access network node, backhaul, and network control programming. Chia et al.~\cite{chia2009next} focused technical challenges associated with the cellular backhaul and presented several existing solutions. The author discussed the current status of mobile backhaul with respect to mobile operators and the near future technologies adopted by them. The authors put forward various suggestions to find cost-effective solutions and technology that will increase the backhaul capacity and efficiency. Tipmongkolsilp et al.~\cite{tipmongkolsilp2011evolution} discussed current issues and technologies of cellular backhaul along with its challenges with respect to the packet switched networks and wireless technologies. The authors also took into account the timing and synchronization issues over the network nodes.

Humair Raza~\cite{raza2011brief} presented an alternative design to counterfeit challenges and issues of backhaul for RANs. Furthermore, Raza and Networks~\cite{raza2013brief} presented backhaul LTE requirements, discussed the issues of synchronization, security, QoS, and path optimization. The authors also mentioned the emerging technologies and showed that throughput directly affects the size and handling capacity of the backhaul devices. The authors discussed the issue of synchronization when the LTE is operated in a time-division duplex mode and then examined the available solutions like ITU standard G.8261-based synchronous Ethernet and IEEE standard 1588v2 based-precision time protocol. Checko et al.~\cite{checko2015cloud} presented state-of-the-art solutions related to C-RAN and its architecture. The authors enumerated the benefits of C-RAN such as to improve the system, mobility, and coverage performance and reduce deployment and operational cost. Ge et al.~\cite{ge20145g} discussed throughput and energy efficiency of 5G wireless backhaul networks emphasizing ultra-dense small cells and mmWave communications. The authors also discussed some future challenges on the adoption of 5G and small cells. Additionally, the authors showed that the deployment of ultra-dense small cells will directly impact the backhaul gateway. Furthermore, it was concluded that the 5G-based handover needs more cooperative small cell group for better backhaul performance.

Wang et al.~\cite{wang2015backhauling} discussed design issues and challenges for radio resource management like cell association, interference management and scheduling in the backhaul. The authors presented a case study of massive Multiple Input Multiple Output (MIMO)-based wireless backhauling for ultra-dense small cells. Various issues like distributed backhaul-aware cell association and inter-cell coordination with heterogeneous backhaul are discussed in detail. Jaber et al.~\cite{jaber20165g} presented a study on 5G backhaul, new trends of backhauling and provided a new 5G backhaul framework. The authors discussed the requirements of 5G backhaul. Various features related to energy, SDN, and mmWave in backhaul were also discussed. Larsen et al.~\cite{larsen2018survey} discussed the current trends of functional splits proposed by 3GPP and other standardization organizations. Santos~\cite{santos20185g} developed an SDN-based architecture for managing small cell backhaul networks. In spite of these attainments, in order to make 5G wireless backhaul networks effective and reliable through a low energy consumption, many factors such as joint cell association, interference management, bandwidth allocation, handover, reducing traffic and synchronization should be considered.

\begin{landscape}
\begin{center}
\fontsize{6}{7}\selectfont
\setlength\LTleft{10pt}            % default: \parindent
\setlength\LTright{0pt}
%\begin{longtable}{llllllllllll}
\begin{longtable}{@{\extracolsep{\fill}}*{15}{c}}
\caption{A comparison of existing surveys and roadmap for mobile backhaul.(R1:Attacks and Threats, R2:Security, R3:5G, R4:SDN, R5:MIMO, R6:mmWave, R7:Microwave, R8:RAN, R9:Handover, R10:Synchronization, R11:OpEx, R12:CapEx, R13:Digital Subscriber Line (DSL))}\label{Table1} \\
\hline
\\
\multicolumn{1}{p{1cm}}{\centering \textbf{Survey}}
&\multicolumn{1}{p{2cm}}{\centering \textbf{Key Contributions}}
&\multicolumn{1}{p{0.5cm}}{\centering \textbf{R1} }
&\multicolumn{1}{p{0.5cm}}{\centering \textbf{R2}}
&\multicolumn{1}{p{0.5cm}}{\centering \textbf{R3}}
&\multicolumn{1}{p{0.5cm}}{\centering \textbf{R4}}
&\multicolumn{1}{p{0.5cm}}{\centering \textbf{R5}}
&\multicolumn{1}{p{0.5cm}}{\centering \textbf{R6}}
&\multicolumn{1}{p{0.5cm}}{\centering \textbf{R7}}
&\multicolumn{1}{p{0.5cm}}{\centering \textbf{R8}}
&\multicolumn{1}{p{0.5cm}}{\centering \textbf{R9}}
&\multicolumn{1}{p{0.5cm}}{\centering \textbf{R10}}
&\multicolumn{1}{p{0.5cm}}{\centering \textbf{R11}}
&\multicolumn{1}{p{0.5cm}}{\centering \textbf{R12}}
&\multicolumn{1}{p{0.5cm}}{\centering \textbf{R13}}
\\[6pt]\\
\hline \\%data entry for gu

\endfirsthead

\multicolumn{15}{c}%
{{\bfseries \tablename\ \thetable{} -- continued from previous page}} \\
\hline
\multicolumn{1}{p{1cm}}{\centering \textbf{Survey}}
&\multicolumn{1}{p{2cm}}{\centering \textbf{Key Contributions}}
&\multicolumn{1}{p{0.5cm}}{\centering \textbf{R1} }
&\multicolumn{1}{p{0.5cm}}{\centering \textbf{R2}}
&\multicolumn{1}{p{0.5cm}}{\centering \textbf{R3}}
&\multicolumn{1}{p{0.5cm}}{\centering \textbf{R4}}
&\multicolumn{1}{p{0.5cm}}{\centering \textbf{R5}}
&\multicolumn{1}{p{0.5cm}}{\centering \textbf{R6}}
&\multicolumn{1}{p{0.5cm}}{\centering \textbf{R7}}
&\multicolumn{1}{p{0.5cm}}{\centering \textbf{R8}}
&\multicolumn{1}{p{0.5cm}}{\centering \textbf{R9}}
&\multicolumn{1}{p{0.5cm}}{\centering \textbf{R10}}
&\multicolumn{1}{p{0.5cm}}{\centering \textbf{R11}}
&\multicolumn{1}{p{0.5cm}}{\centering \textbf{R12}}
&\multicolumn{1}{p{0.5cm}}{\centering \textbf{R13}}
\\[6pt]\\
\hline\\\endhead

\hline \multicolumn{15}{l}{{Continued on next page}} \\

\endfoot

\endlastfoot

\multicolumn{1}{p{1cm}}{\centering Saha et al. \cite{saha2016evolution}}
&\multicolumn{1}{p{2cm}}{\centering Presented discussion on backhaul networks and network synchronization in 5G mobile networks}
&\multicolumn{1}{p{0.5cm}}{\centering \xmark}
&\multicolumn{1}{p{0.5cm}}{\centering \xmark}
&\multicolumn{1}{p{0.5cm}}{\centering  \cmark}
&\multicolumn{1}{p{0.5cm}}{\centering \cmark}
&\multicolumn{1}{p{0.5cm}}{\centering \cmark}
&\multicolumn{1}{p{0.5cm}}{\centering \cmark}
&\multicolumn{1}{p{0.5cm}}{\centering \cmark}
&\multicolumn{1}{p{0.5cm}}{\centering \cmark}
&\multicolumn{1}{p{0.5cm}}{\centering  \cmark}
&\multicolumn{1}{p{0.5cm}}{\centering \cmark}
&\multicolumn{1}{p{0.5cm}}{\centering \xmark}
&\multicolumn{1}{p{0.5cm}}{\centering  \xmark}
&\multicolumn{1}{p{0.5cm}}{\centering \cmark}   \\[6pt]\\

\multicolumn{1}{p{1cm}}{\centering Chia et al. \cite{chia2009next}}
&\multicolumn{1}{p{2.5cm}}{\centering Discussed the technical
challenges faced by wireless operators and technical solutions}
&\multicolumn{1}{p{0.5cm}}{\centering \xmark}
&\multicolumn{1}{p{0.5cm}}{\centering \xmark}
&\multicolumn{1}{p{0.5cm}}{\centering  \xmark}
&\multicolumn{1}{p{0.5cm}}{\centering \xmark}
&\multicolumn{1}{p{0.5cm}}{\centering \xmark}
&\multicolumn{1}{p{0.5cm}}{\centering \cmark}
&\multicolumn{1}{p{0.5cm}}{\centering \cmark}
&\multicolumn{1}{p{0.5cm}}{\centering \cmark}
&\multicolumn{1}{p{0.5cm}}{\centering  \cmark}
&\multicolumn{1}{p{0.5cm}}{\centering \cmark}
&\multicolumn{1}{p{0.5cm}}{\centering \xmark}
&\multicolumn{1}{p{0.5cm}}{\centering  \xmark}
&\multicolumn{1}{p{0.5cm}}{\centering \cmark}   \\[6pt]\\

\multicolumn{1}{p{1cm}}{\centering Tipmongkolsilp et al. \cite{tipmongkolsilp2011evolution}}
&\multicolumn{1}{p{2.5cm}}{\centering Discussed current issues, technologies, and challenges of cellular backhaul.}
&\multicolumn{1}{p{0.5cm}}{\centering \xmark}
&\multicolumn{1}{p{0.5cm}}{\centering \cmark}
&\multicolumn{1}{p{0.5cm}}{\centering  \xmark}
&\multicolumn{1}{p{0.5cm}}{\centering \xmark}
&\multicolumn{1}{p{0.5cm}}{\centering \cmark}
&\multicolumn{1}{p{0.5cm}}{\centering \xmark}
&\multicolumn{1}{p{0.5cm}}{\centering \cmark}
&\multicolumn{1}{p{0.5cm}}{\centering \xmark}
&\multicolumn{1}{p{0.5cm}}{\centering  \cmark}
&\multicolumn{1}{p{0.5cm}}{\centering \cmark}
&\multicolumn{1}{p{0.5cm}}{\centering \cmark}
&\multicolumn{1}{p{0.5cm}}{\centering  \cmark}
&\multicolumn{1}{p{0.5cm}}{\centering \cmark}   \\[6pt]\\

\multicolumn{1}{p{1cm}}{\centering Humair Raza\cite{raza2011brief}}
&\multicolumn{1}{p{2.5cm}}{\centering Presented evolution of RAN backhaul technologies}
&\multicolumn{1}{p{0.5cm}}{\centering \xmark}
&\multicolumn{1}{p{0.5cm}}{\centering \xmark}
&\multicolumn{1}{p{0.5cm}}{\centering  \xmark}
&\multicolumn{1}{p{0.5cm}}{\centering \xmark}
&\multicolumn{1}{p{0.5cm}}{\centering \xmark}
&\multicolumn{1}{p{0.5cm}}{\centering \xmark}
&\multicolumn{1}{p{0.5cm}}{\centering \cmark}
&\multicolumn{1}{p{0.5cm}}{\centering \cmark}
&\multicolumn{1}{p{0.5cm}}{\centering  \xmark}
&\multicolumn{1}{p{0.5cm}}{\centering \xmark}
&\multicolumn{1}{p{0.5cm}}{\centering \xmark}
&\multicolumn{1}{p{0.5cm}}{\centering  \xmark}
&\multicolumn{1}{p{0.5cm}}{\centering \cmark}   \\[6pt]\\

\multicolumn{1}{p{1cm}}{\centering Raza and Networks\cite{raza2013brief}}
&\multicolumn{1}{p{2.5cm}}{\centering Discussed backhaul technology for LTE-based RANs}
&\multicolumn{1}{p{0.5cm}}{\centering \xmark}
&\multicolumn{1}{p{0.5cm}}{\centering \cmark}
&\multicolumn{1}{p{0.5cm}}{\centering  \xmark}
&\multicolumn{1}{p{0.5cm}}{\centering \xmark}
&\multicolumn{1}{p{0.5cm}}{\centering \cmark}
&\multicolumn{1}{p{0.5cm}}{\centering \xmark}
&\multicolumn{1}{p{0.5cm}}{\centering \xmark}
&\multicolumn{1}{p{0.5cm}}{\centering \cmark}
&\multicolumn{1}{p{0.5cm}}{\centering  \cmark}
&\multicolumn{1}{p{0.5cm}}{\centering \cmark}
&\multicolumn{1}{p{0.5cm}}{\centering \xmark}
&\multicolumn{1}{p{0.5cm}}{\centering  \xmark}
&\multicolumn{1}{p{0.5cm}}{\centering \cmark}   \\[6pt]\\

\multicolumn{1}{p{1cm}}{\centering Checko et al. \cite{checko2015cloud}}
&\multicolumn{1}{p{2.5cm}}{\centering Presented the technological aspects of C-RAN and the related challenges.}
&\multicolumn{1}{p{0.5cm}}{\centering \xmark}
&\multicolumn{1}{p{0.5cm}}{\centering \cmark}
&\multicolumn{1}{p{0.5cm}}{\centering  \cmark}
&\multicolumn{1}{p{0.5cm}}{\centering \cmark}
&\multicolumn{1}{p{0.5cm}}{\centering \cmark}
&\multicolumn{1}{p{0.5cm}}{\centering \xmark}
&\multicolumn{1}{p{0.5cm}}{\centering \cmark}
&\multicolumn{1}{p{0.5cm}}{\centering \cmark}
&\multicolumn{1}{p{0.5cm}}{\centering  \cmark}
&\multicolumn{1}{p{0.5cm}}{\centering \cmark}
&\multicolumn{1}{p{0.5cm}}{\centering \cmark}
&\multicolumn{1}{p{0.5cm}}{\centering  \cmark}
&\multicolumn{1}{p{0.5cm}}{\centering \cmark}   \\[6pt]\\

\multicolumn{1}{p{1cm}}{\centering Ge et al. \cite{ge20145g}}
&\multicolumn{1}{p{2.5cm}}{\centering Discussed throughput and energy efficiency of 5G wireless backhaul networks and emphasized ultra-dense small cells and mmWave communications.}
&\multicolumn{1}{p{0.5cm}}{\centering \xmark}
&\multicolumn{1}{p{0.5cm}}{\centering \xmark}
&\multicolumn{1}{p{0.5cm}}{\centering  \cmark}
&\multicolumn{1}{p{0.5cm}}{\centering \xmark}
&\multicolumn{1}{p{0.5cm}}{\centering \cmark}
&\multicolumn{1}{p{0.5cm}}{\centering \cmark}
&\multicolumn{1}{p{0.5cm}}{\centering \cmark}
&\multicolumn{1}{p{0.5cm}}{\centering \xmark}
&\multicolumn{1}{p{0.5cm}}{\centering  \cmark}
&\multicolumn{1}{p{0.5cm}}{\centering \xmark}
&\multicolumn{1}{p{0.5cm}}{\centering \xmark}
&\multicolumn{1}{p{0.5cm}}{\centering  \xmark}
&\multicolumn{1}{p{0.5cm}}{\centering \xmark}   \\[6pt]\\

\multicolumn{1}{p{1cm}}{\centering Wang et al. \cite{wang2015backhauling}}
&\multicolumn{1}{p{2.5cm}}{\centering Discussed design issues
and challenges for radio resource management with backhaul}
&\multicolumn{1}{p{0.5cm}}{\centering \xmark}
&\multicolumn{1}{p{0.5cm}}{\centering \xmark}
&\multicolumn{1}{p{0.5cm}}{\centering  \cmark}
&\multicolumn{1}{p{0.5cm}}{\centering \xmark}
&\multicolumn{1}{p{0.5cm}}{\centering \cmark}
&\multicolumn{1}{p{0.5cm}}{\centering \cmark}
&\multicolumn{1}{p{0.5cm}}{\centering \cmark}
&\multicolumn{1}{p{0.5cm}}{\centering \cmark}
&\multicolumn{1}{p{0.5cm}}{\centering  \cmark}
&\multicolumn{1}{p{0.5cm}}{\centering \xmark}
&\multicolumn{1}{p{0.5cm}}{\centering \xmark}
&\multicolumn{1}{p{0.5cm}}{\centering  \xmark}
&\multicolumn{1}{p{0.5cm}}{\centering \cmark}   \\[6pt]\\

\multicolumn{1}{p{1cm}}{\centering Jaber et al. \cite{jaber20165g}}
&\multicolumn{1}{p{2.5cm}}{\centering Presented study on 5G backhaul, new trends of backhauling and gave a new 5G backhaul framework}
&\multicolumn{1}{p{0.5cm}}{\centering \xmark}
&\multicolumn{1}{p{0.5cm}}{\centering \cmark}
&\multicolumn{1}{p{0.5cm}}{\centering  \cmark}
&\multicolumn{1}{p{0.5cm}}{\centering \cmark}
&\multicolumn{1}{p{0.5cm}}{\centering \cmark}
&\multicolumn{1}{p{0.5cm}}{\centering \cmark}
&\multicolumn{1}{p{0.5cm}}{\centering \cmark}
&\multicolumn{1}{p{0.5cm}}{\centering \cmark}
&\multicolumn{1}{p{0.5cm}}{\centering  \cmark}
&\multicolumn{1}{p{0.5cm}}{\centering \cmark}
&\multicolumn{1}{p{0.5cm}}{\centering \cmark}
&\multicolumn{1}{p{0.5cm}}{\centering  \cmark}
&\multicolumn{1}{p{0.5cm}}{\centering \cmark}   \\[6pt]\\

\multicolumn{1}{p{1cm}}{\centering Larsen et al. \cite{larsen2018survey}}
&\multicolumn{1}{p{2.5cm}}{\centering Discussed the current trend of functional splits proposed by 3GPP and other standardization organizations}
&\multicolumn{1}{p{0.5cm}}{\centering \xmark}
&\multicolumn{1}{p{0.5cm}}{\centering \cmark}
&\multicolumn{1}{p{0.5cm}}{\centering  \cmark}
&\multicolumn{1}{p{0.5cm}}{\centering \xmark}
&\multicolumn{1}{p{0.5cm}}{\centering \cmark}
&\multicolumn{1}{p{0.5cm}}{\centering \xmark}
&\multicolumn{1}{p{0.5cm}}{\centering \cmark}
&\multicolumn{1}{p{0.5cm}}{\centering \cmark}
&\multicolumn{1}{p{0.5cm}}{\centering  \cmark}
&\multicolumn{1}{p{0.5cm}}{\centering \cmark}
&\multicolumn{1}{p{0.5cm}}{\centering \xmark}
&\multicolumn{1}{p{0.5cm}}{\centering  \xmark}
&\multicolumn{1}{p{0.5cm}}{\centering \xmark}   \\[6pt]\\

\multicolumn{1}{p{1cm}}{\centering Ricardo Santos\cite{santos20185g}}
&\multicolumn{1}{p{2.5cm}}{\centering Proposed an SDN-based architecture for managing small cell backhaul networks}
&\multicolumn{1}{p{0.5cm}}{\centering \xmark}
&\multicolumn{1}{p{0.5cm}}{\centering \xmark}
&\multicolumn{1}{p{0.5cm}}{\centering  \cmark}
&\multicolumn{1}{p{0.5cm}}{\centering \cmark}
&\multicolumn{1}{p{0.5cm}}{\centering \cmark}
&\multicolumn{1}{p{0.5cm}}{\centering \cmark}
&\multicolumn{1}{p{0.5cm}}{\centering \cmark}
&\multicolumn{1}{p{0.5cm}}{\centering \xmark}
&\multicolumn{1}{p{0.5cm}}{\centering  \xmark}
&\multicolumn{1}{p{0.5cm}}{\centering \xmark}
&\multicolumn{1}{p{0.5cm}}{\centering \xmark}
&\multicolumn{1}{p{0.5cm}}{\centering  \xmark}
&\multicolumn{1}{p{0.5cm}}{\centering \xmark}   \\[6pt]\\
\hline

\end{longtable}%}
\end{center}
\end{landscape}

\section{5G Mobile Backhaul Networks}
The communication network comprises various modules, which play a significant role in data transmission. The new architectures of mobile backhaul are coped up with the emerging demands of the 5G networks, fast data rates, and low latency~\cite{asif20185g}. The mobile backhaul operators, new competitors, and technologies are improving average revenue per user while trying to reduce the operational and deployment costs. The traditional bandwidth facilitates voice calls, SMS, etc, but the emerging demands of users require various services like real-time gaming, video streaming, video conferencing and calling which necessitate a high bandwidth with ultra-low latency~\cite{dighriri2018measurement}.

The traditional operator categorized networks according to the requirements and builds multiple sub networks. Nowadays, the operators are focusing on the optimized use of a single network in multiple services as well as backhaul. The new mobile backhaul is moving from traditional technologies such as Time-Division Multiplexing (TDM), ATM, and Internet Protocol (IP)/Ethernet-based to technologies such as WiMAX- Long-Term Evolution (LTE), and 5G backhaul~\cite{chapman2018low}. The mobile backhaul operators emphasized various key factors such as VPN-based services across the network, User traffic prediction, QoS, and high capacity with scalable and cost-effective setups to reduce the bottleneck problems. The need for a backhaul solution which can efficiently and affordably manage the data is increased especially due to the requirements of greater and more sophisticated high-speed wireless data services for the customers.

\subsection{Categorization of 5G Mobile Backhaul Networks}
\begin{figure}[ht!]
\centering
\includegraphics[width=400px]{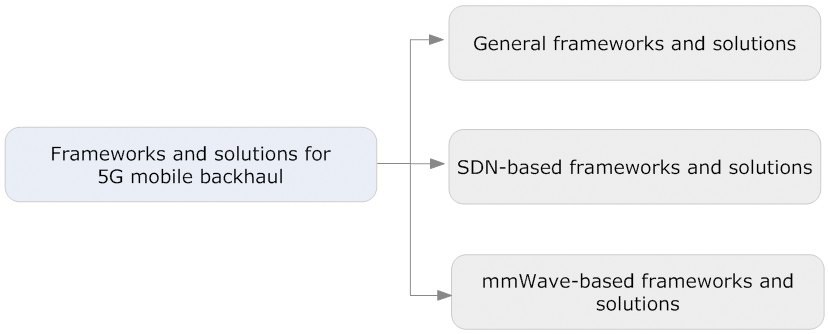}
\caption{A taxonomy of existing frameworks and solutions for 5G mobile backhaul.}
\label{tax}
\end{figure}

The new technology adoption in the era of mobile communication involves various add-ons in the network and devices. The backhaul plays a significant role in the connectivity of the base stations and radio controllers. The increasing demands for capacity and traffic management lead to fast backhaul requirements~\cite{larsen2018survey}. The add-ons in the technology have huge impacts on the cost and deployment of the new devices and policies. The emerging backhaul network technologies and solutions have been accentuated in various research articles. This survey examines the existing 5G backhaul solutions and architectures from the academic as well as industrial sectors. The taxonomy, as shown in Fig\ref{tax}, of existing 5G mobile backhaul framework is categorized into three major types, namely, general frameworks and solutions, SDN- based frameworks and solutions, and mmWave-based frameworks and solutions.

SDN and mmWave-based backhaul solutions can be considered as effective strategies which lead to a better compatibility among users, services and service providers. These solutions reduce the capital cost and improve the performance and energy efficiency of the networks.

\subsubsection{General Frameworks and Solutions}
In the general framework classifications of 5G, mobile backhaul consists of backhaul management and efficient data offloading schemes. In the 5G heterogeneous network, the backhaul management is a challenging task due to massive data inputs and outputs. Some researchers emphasized the energy efficient smart backhauling. Huskov et al.\cite{huskov2015smart} focused on the smart backhauling and designed a new backhauling subsystem based on the smart resource allocations in the small cells. The proposed solution relies on the assumptions of conditional operability of the network components. The conditions are defined with the significance and complexities. The concept of multiple input distributed output (MIDO) technology is used in the backhauling subsystem. The proposed smart backhauling solution shows energy and spectral efficiency. The authors in \cite{beyranvand2017toward} discussed the 5G key issues along with backhaul reliability concerns from passive optical network fiber faults. Consequently, they provided a capacity centric fiber-wireless enhanced LTE-A Heterogeneous networks (HetNets) architecture and proposed DOFR routing algorithm which helps in the improvement of aggregate throughput. Hamidouche et al.\cite{hamidouche20175g} presented the caching solutions for small cell networks and employed distributed backhaul management approach. In the proposed scheme, the backhaul management problem is formulated as a Minority Game. The authors further proposed self-organizing reinforcement learning algorithm as a progressive solution.

Variety of solutions has emphasized the efficient resource allocation strategies. The MAC (Medium Access Control) solutions have also been described in the 5G backhaul. Krasko et al.\cite{krasko2017enhanced} explained the resource allocation of 5G mobile backhaul and illustrated an enhanced MAC strategy. The proposed MAC provides compatibility among the optical backhaul and wireless backhaul. On the other hand, Alzenad et al.\cite{alzenad2018fso} proposed a free space optics-based backhaul framework for high data rates. The author also considered factors which affect the Free-Space Optical (FSO) signals like absorption, scattering, and turbulence. The CapEx and OpEx cost were also discussed in the adoption of FSO links. Furthermore, various challenges associated with the deployment of FSO architecture were illustrated. Massive multiple input massive multiple output data issues were addressed by Huy et al.\cite{phan2017massive}. The authors showed a signal processing scheme for massive multiple inputs and massive multiple outputs by considering the issues related to low complexity.

Jaber et al.\cite{jaber2018wireless} emphasized the wireless backhaul and presented a performance analyzing framework in the Line of Sight (LOS) environment. The authors considered the throughput, latency, and resilience factors in the wireless backhaul. Atakora et al.\cite{atakora2018multicast} focused on the optimal multicast in mobile scenarios in 5G backhaul networks and illustrated the multicast technique. The authors mainly focused on the physical layer multicasting mobility and formulated it into traveling salesmen problem. Energy efficient solutions for passive optical network and mmWave backhaul technologies are also presented by Mowla et al.\cite{mowla2018energy}. The authors formulated the small cell network problem as a min-cost network flow optimization problem. Maksymyuk et al.\cite{maksymyuk2017designing} developed new backhaul design on the basis of time-frequency resource grid. To reduce the handover overheads, a new handover mechanism has been designed based on their proposed backhaul design. Haddaji et al.\cite{haddaji2018backhauling} illustrated a cost of ownership backHauling-as-a-Service analysis method to improve the performance and accuracy of network planning.
The existing solutions considered the parameters like reliability, flexibility, QoS without compromising the performance factors. The existing general frameworks and solutions for 5G mobile backhaul are presented in Table-\ref{Tablegeneral}.

\begin{landscape}
\begin{center}
\fontsize{6}{6}\selectfont
\setlength\LTleft{10pt}            % default: \parindent
\setlength\LTright{0pt}
%\begin{longtable}{llllllllllll}
\begin{longtable}{@{\extracolsep{\fill}}*{19}{c}}
\caption{A comparison of existing general frameworks and solutions for 5G mobile backhaul networks.(R1:Latency/Delay Consideration, R2:Capacity Enhancement, R3:Quality of
Service, R4:Easy Installation/Cost Effectiveness, R5:Traffic Management, R6:Synchronization, R7:Energy Considerations, R8:Complexity Consideration, R9:Fronthaul, R10:Multi-tenancy/Network Slicing support, R11:Handover Support, R12:Link Utilization, R13:Load Balancing, R14:RAN, R15:SON, R16:MAC Protocol Considerations, R17:Considerations of Standards )}\label{Tablegeneral} \\
\hline
\\[6pt]\\
\multicolumn{1}{p{1cm}}{\centering \textbf{Scheme}}
&\multicolumn{1}{p{2cm}}{\centering \textbf{Main Contributions}}
&\multicolumn{1}{p{0.5cm}}{\centering \textbf{R1} }
&\multicolumn{1}{p{0.5cm}}{\centering \textbf{R2}}
&\multicolumn{1}{p{0.5cm}}{\centering \textbf{R3}}
&\multicolumn{1}{p{0.5cm}}{\centering \textbf{R4}}
&\multicolumn{1}{p{0.5cm}}{\centering \textbf{R5}}
&\multicolumn{1}{p{0.5cm}}{\centering \textbf{R6}}
&\multicolumn{1}{p{0.5cm}}{\centering \textbf{R7}}
&\multicolumn{1}{p{0.5cm}}{\centering \textbf{R8}}
&\multicolumn{1}{p{0.5cm}}{\centering \textbf{R9}}
&\multicolumn{1}{p{0.5cm}}{\centering \textbf{R10}}
&\multicolumn{1}{p{0.5cm}}{\centering \textbf{R11}}
&\multicolumn{1}{p{0.5cm}}{\centering \textbf{R12}}
&\multicolumn{1}{p{0.5cm}}{\centering \textbf{R13}}
&\multicolumn{1}{p{0.5cm}}{\centering \textbf{R14}}
&\multicolumn{1}{p{0.5cm}}{\centering \textbf{R15}}
&\multicolumn{1}{p{0.5cm}}{\centering \textbf{R16}}
&\multicolumn{1}{p{0.5cm}}{\centering \textbf{R17}}
\\[6pt]\\
\hline \\%data entry for gu

\endfirsthead

\multicolumn{19}{c}%
{{\bfseries \tablename\ \thetable{} -- continued from previous page}} \\
\hline
\multicolumn{1}{p{1cm}}{\centering \textbf{Scheme}}
&\multicolumn{1}{p{2cm}}{\centering \textbf{Main Contributions}}
&\multicolumn{1}{p{0.5cm}}{\centering \textbf{R1} }
&\multicolumn{1}{p{0.5cm}}{\centering \textbf{R2}}
&\multicolumn{1}{p{0.5cm}}{\centering \textbf{R3}}
&\multicolumn{1}{p{0.5cm}}{\centering \textbf{R4}}
&\multicolumn{1}{p{0.5cm}}{\centering \textbf{R5}}
&\multicolumn{1}{p{0.5cm}}{\centering \textbf{R6}}
&\multicolumn{1}{p{0.5cm}}{\centering \textbf{R7}}
&\multicolumn{1}{p{0.5cm}}{\centering \textbf{R8}}
&\multicolumn{1}{p{0.5cm}}{\centering \textbf{R9}}
&\multicolumn{1}{p{0.5cm}}{\centering \textbf{R10}}
&\multicolumn{1}{p{0.5cm}}{\centering \textbf{R11}}
&\multicolumn{1}{p{0.5cm}}{\centering \textbf{R12}}
&\multicolumn{1}{p{0.5cm}}{\centering \textbf{R13}}
&\multicolumn{1}{p{0.5cm}}{\centering \textbf{R14}}
&\multicolumn{1}{p{0.5cm}}{\centering \textbf{R15}}
&\multicolumn{1}{p{0.5cm}}{\centering \textbf{R16}}
&\multicolumn{1}{p{0.5cm}}{\centering \textbf{R17}}
\\[6pt]\\
\hline \\
\endhead

\hline \multicolumn{19}{l}{{Continued on next page}} \\

\endfoot

\endlastfoot

\multicolumn{1}{p{1cm}}{\centering Huskov et al.\cite{huskov2015smart}}
&\multicolumn{1}{p{2cm}}{\centering Proposed a new backhauling subsystem based on the smart resource allocation}
&\multicolumn{1}{p{0.5cm}}{\centering \xmark}
&\multicolumn{1}{p{0.5cm}}{\centering \cmark}
&\multicolumn{1}{p{0.5cm}}{\centering  \xmark}
&\multicolumn{1}{p{0.5cm}}{\centering \cmark}
&\multicolumn{1}{p{0.5cm}}{\centering \xmark}
&\multicolumn{1}{p{0.5cm}}{\centering \xmark}
&\multicolumn{1}{p{0.5cm}}{\centering \cmark}
&\multicolumn{1}{p{0.5cm}}{\centering \cmark}
&\multicolumn{1}{p{0.5cm}}{\centering  \xmark}
&\multicolumn{1}{p{0.5cm}}{\centering \xmark}
&\multicolumn{1}{p{0.5cm}}{\centering \xmark}
&\multicolumn{1}{p{0.5cm}}{\centering \cmark}
&\multicolumn{1}{p{0.5cm}}{\centering \cmark}
&\multicolumn{1}{p{0.5cm}}{\centering \xmark}
&\multicolumn{1}{p{0.5cm}}{\centering  \xmark}
&\multicolumn{1}{p{0.5cm}}{\centering \xmark}
&\multicolumn{1}{p{0.5cm}}{\centering \xmark}
\\[6pt]\\

\multicolumn{1}{p{1cm}}{\centering Beyranvand et al.\cite{beyranvand2017toward}}
&\multicolumn{1}{p{2cm}}{\centering Described the backhaul reliability issues stemming from passive optical network fiber faults}
&\multicolumn{1}{p{0.5cm}}{\centering \cmark}
&\multicolumn{1}{p{0.5cm}}{\centering \cmark}
&\multicolumn{1}{p{0.5cm}}{\centering  \xmark}
&\multicolumn{1}{p{0.5cm}}{\centering \xmark}
&\multicolumn{1}{p{0.5cm}}{\centering \cmark}
&\multicolumn{1}{p{0.5cm}}{\centering \xmark}
&\multicolumn{1}{p{0.5cm}}{\centering \xmark}
&\multicolumn{1}{p{0.5cm}}{\centering \xmark}
&\multicolumn{1}{p{0.5cm}}{\centering  \xmark}
&\multicolumn{1}{p{0.5cm}}{\centering \xmark}
&\multicolumn{1}{p{0.5cm}}{\centering \xmark}
&\multicolumn{1}{p{0.5cm}}{\centering \cmark}
&\multicolumn{1}{p{0.5cm}}{\centering -}
&\multicolumn{1}{p{0.5cm}}{\centering \xmark}
&\multicolumn{1}{p{0.5cm}}{\centering  \xmark}
&\multicolumn{1}{p{0.5cm}}{\centering -}
&\multicolumn{1}{p{0.5cm}}{\centering \cmark}
\\[6pt]\\

\multicolumn{1}{p{1cm}}{\centering Hamidouche et al.\cite{hamidouche20175g}}
&\multicolumn{1}{p{2cm}}{\centering Proposed a distributed backhaul management approach}
&\multicolumn{1}{p{0.5cm}}{\centering \cmark}
&\multicolumn{1}{p{0.5cm}}{\centering \cmark}
&\multicolumn{1}{p{0.5cm}}{\centering  \cmark}
&\multicolumn{1}{p{0.5cm}}{\centering -}
&\multicolumn{1}{p{0.5cm}}{\centering \cmark}
&\multicolumn{1}{p{0.5cm}}{\centering \xmark}
&\multicolumn{1}{p{0.5cm}}{\centering \xmark}
&\multicolumn{1}{p{0.5cm}}{\centering -}
&\multicolumn{1}{p{0.5cm}}{\centering  \xmark}
&\multicolumn{1}{p{0.5cm}}{\centering \xmark}
&\multicolumn{1}{p{0.5cm}}{\centering \xmark}
&\multicolumn{1}{p{0.5cm}}{\centering \cmark}
&\multicolumn{1}{p{0.5cm}}{\centering \cmark}
&\multicolumn{1}{p{0.5cm}}{\centering \xmark}
&\multicolumn{1}{p{0.5cm}}{\centering  \xmark}
&\multicolumn{1}{p{0.5cm}}{\centering \xmark}
&\multicolumn{1}{p{0.5cm}}{\centering \cmark}
\\[6pt]\\

\multicolumn{1}{p{1cm}}{\centering Krasko et al.\cite{krasko2017enhanced}}
&\multicolumn{1}{p{2cm}}{\centering Focused on the resource allocation of 5G mobile backhaul and gives an enhanced MAC}
&\multicolumn{1}{p{0.5cm}}{\centering \xmark}
&\multicolumn{1}{p{0.5cm}}{\centering \cmark}
&\multicolumn{1}{p{0.5cm}}{\centering  \xmark}
&\multicolumn{1}{p{0.5cm}}{\centering \cmark}
&\multicolumn{1}{p{0.5cm}}{\centering -}
&\multicolumn{1}{p{0.5cm}}{\centering \xmark}
&\multicolumn{1}{p{0.5cm}}{\centering \xmark}
&\multicolumn{1}{p{0.5cm}}{\centering \xmark}
&\multicolumn{1}{p{0.5cm}}{\centering  \xmark}
&\multicolumn{1}{p{0.5cm}}{\centering \xmark}
&\multicolumn{1}{p{0.5cm}}{\centering \xmark}
&\multicolumn{1}{p{0.5cm}}{\centering \xmark}
&\multicolumn{1}{p{0.5cm}}{\centering \xmark}
&\multicolumn{1}{p{0.5cm}}{\centering \cmark}
&\multicolumn{1}{p{0.5cm}}{\centering  \xmark}
&\multicolumn{1}{p{0.5cm}}{\centering \cmark}
&\multicolumn{1}{p{0.5cm}}{\centering -}
\\[6pt]\\

\multicolumn{1}{p{1cm}}{\centering Alzenad et al.\cite{alzenad2018fso}}
&\multicolumn{1}{p{2cm}}{\centering Proposed a FSO-based backhaul framework for high data rates}
&\multicolumn{1}{p{0.5cm}}{\centering \xmark}
&\multicolumn{1}{p{0.5cm}}{\centering -}
&\multicolumn{1}{p{0.5cm}}{\centering  \xmark}
&\multicolumn{1}{p{0.5cm}}{\centering \cmark}
&\multicolumn{1}{p{0.5cm}}{\centering \cmark}
&\multicolumn{1}{p{0.5cm}}{\centering \xmark}
&\multicolumn{1}{p{0.5cm}}{\centering \cmark}
&\multicolumn{1}{p{0.5cm}}{\centering \xmark}
&\multicolumn{1}{p{0.5cm}}{\centering  \cmark}
&\multicolumn{1}{p{0.5cm}}{\centering \xmark}
&\multicolumn{1}{p{0.5cm}}{\centering \xmark}
&\multicolumn{1}{p{0.5cm}}{\centering \cmark}
&\multicolumn{1}{p{0.5cm}}{\centering -}
&\multicolumn{1}{p{0.5cm}}{\centering \cmark}
&\multicolumn{1}{p{0.5cm}}{\centering  \xmark}
&\multicolumn{1}{p{0.5cm}}{\centering \xmark}
&\multicolumn{1}{p{0.5cm}}{\centering \cmark}
\\[6pt]\\

\multicolumn{1}{p{1cm}}{\centering Huy et al.\cite{phan2017massive}}
&\multicolumn{1}{p{2cm}}{\centering Proposed a signal
processing scheme for massive multiple input massive multiple output}
&\multicolumn{1}{p{0.5cm}}{\centering \xmark}
&\multicolumn{1}{p{0.5cm}}{\centering \xmark}
&\multicolumn{1}{p{0.5cm}}{\centering  \xmark}
&\multicolumn{1}{p{0.5cm}}{\centering \xmark}
&\multicolumn{1}{p{0.5cm}}{\centering \xmark}
&\multicolumn{1}{p{0.5cm}}{\centering \xmark}
&\multicolumn{1}{p{0.5cm}}{\centering \cmark}
&\multicolumn{1}{p{0.5cm}}{\centering \cmark}
&\multicolumn{1}{p{0.5cm}}{\centering  \xmark}
&\multicolumn{1}{p{0.5cm}}{\centering \xmark}
&\multicolumn{1}{p{0.5cm}}{\centering \xmark}
&\multicolumn{1}{p{0.5cm}}{\centering \cmark}
&\multicolumn{1}{p{0.5cm}}{\centering \xmark}
&\multicolumn{1}{p{0.5cm}}{\centering \xmark}
&\multicolumn{1}{p{0.5cm}}{\centering  \xmark}
&\multicolumn{1}{p{0.5cm}}{\centering \xmark}
&\multicolumn{1}{p{0.5cm}}{\centering \xmark}
\\[6pt]\\

\multicolumn{1}{p{1cm}}{\centering Jaber et al.\cite{jaber2018wireless}}
&\multicolumn{1}{p{2cm}}{\centering Emphasized on the wireless backhaul and presented a performance analyzing framework}
&\multicolumn{1}{p{0.5cm}}{\centering \cmark}
&\multicolumn{1}{p{0.5cm}}{\centering \cmark}
&\multicolumn{1}{p{0.5cm}}{\centering  \cmark}
&\multicolumn{1}{p{0.5cm}}{\centering -}
&\multicolumn{1}{p{0.5cm}}{\centering \xmark}
&\multicolumn{1}{p{0.5cm}}{\centering \xmark}
&\multicolumn{1}{p{0.5cm}}{\centering \xmark}
&\multicolumn{1}{p{0.5cm}}{\centering \cmark}
&\multicolumn{1}{p{0.5cm}}{\centering  -}
&\multicolumn{1}{p{0.5cm}}{\centering \xmark}
&\multicolumn{1}{p{0.5cm}}{\centering \xmark}
&\multicolumn{1}{p{0.5cm}}{\centering \cmark}
&\multicolumn{1}{p{0.5cm}}{\centering \cmark}
&\multicolumn{1}{p{0.5cm}}{\centering \xmark}
&\multicolumn{1}{p{0.5cm}}{\centering  \xmark}
&\multicolumn{1}{p{0.5cm}}{\centering \xmark}
&\multicolumn{1}{p{0.5cm}}{\centering \xmark}
\\[6pt]\\

\multicolumn{1}{p{1cm}}{\centering Atakora et al.\cite{atakora2018multicast}}
&\multicolumn{1}{p{2cm}}{\centering Focused on the optimal multicast in mobile scenarios and gave multicast technique}
&\multicolumn{1}{p{0.5cm}}{\centering \cmark}
&\multicolumn{1}{p{0.5cm}}{\centering \cmark}
&\multicolumn{1}{p{0.5cm}}{\centering  \xmark}
&\multicolumn{1}{p{0.5cm}}{\centering \cmark}
&\multicolumn{1}{p{0.5cm}}{\centering \xmark}
&\multicolumn{1}{p{0.5cm}}{\centering \xmark}
&\multicolumn{1}{p{0.5cm}}{\centering \cmark}
&\multicolumn{1}{p{0.5cm}}{\centering \cmark}
&\multicolumn{1}{p{0.5cm}}{\centering  \xmark}
&\multicolumn{1}{p{0.5cm}}{\centering \xmark}
&\multicolumn{1}{p{0.5cm}}{\centering \xmark}
&\multicolumn{1}{p{0.5cm}}{\centering \cmark}
&\multicolumn{1}{p{0.5cm}}{\centering \xmark}
&\multicolumn{1}{p{0.5cm}}{\centering \xmark}
&\multicolumn{1}{p{0.5cm}}{\centering  \xmark}
&\multicolumn{1}{p{0.5cm}}{\centering -}
&\multicolumn{1}{p{0.5cm}}{\centering \cmark}
\\[6pt]\\

\multicolumn{1}{p{1cm}}{\centering Mowla et al.\cite{mowla2018energy}}
&\multicolumn{1}{p{2cm}}{\centering
Focused on the energy efficient solutions for passive optical network and mmWave backhaul technologies}
&\multicolumn{1}{p{0.5cm}}{\centering \cmark}
&\multicolumn{1}{p{0.5cm}}{\centering \cmark}
&\multicolumn{1}{p{0.5cm}}{\centering  \cmark}
&\multicolumn{1}{p{0.5cm}}{\centering \cmark}
&\multicolumn{1}{p{0.5cm}}{\centering \cmark}
&\multicolumn{1}{p{0.5cm}}{\centering \xmark}
&\multicolumn{1}{p{0.5cm}}{\centering \cmark}
&\multicolumn{1}{p{0.5cm}}{\centering \cmark}
&\multicolumn{1}{p{0.5cm}}{\centering  \xmark}
&\multicolumn{1}{p{0.5cm}}{\centering \xmark}
&\multicolumn{1}{p{0.5cm}}{\centering -}
&\multicolumn{1}{p{0.5cm}}{\centering \cmark}
&\multicolumn{1}{p{0.5cm}}{\centering \cmark}
&\multicolumn{1}{p{0.5cm}}{\centering \cmark}
&\multicolumn{1}{p{0.5cm}}{\centering  \xmark}
&\multicolumn{1}{p{0.5cm}}{\centering \xmark}
&\multicolumn{1}{p{0.5cm}}{\centering -}
\\[6pt]\\

\multicolumn{1}{p{1cm}}{\centering Maksymyuk et al.\cite{maksymyuk2017designing}}
&\multicolumn{1}{p{2cm}}{\centering On the basis of time-frequency resource grid a new backhaul design is presented}
&\multicolumn{1}{p{0.5cm}}{\centering \xmark}
&\multicolumn{1}{p{0.5cm}}{\centering \cmark}
&\multicolumn{1}{p{0.5cm}}{\centering  \xmark}
&\multicolumn{1}{p{0.5cm}}{\centering \xmark}
&\multicolumn{1}{p{0.5cm}}{\centering \cmark}
&\multicolumn{1}{p{0.5cm}}{\centering \xmark}
&\multicolumn{1}{p{0.5cm}}{\centering -}
&\multicolumn{1}{p{0.5cm}}{\centering \xmark}
&\multicolumn{1}{p{0.5cm}}{\centering  \cmark}
&\multicolumn{1}{p{0.5cm}}{\centering \xmark}
&\multicolumn{1}{p{0.5cm}}{\centering \cmark}
&\multicolumn{1}{p{0.5cm}}{\centering \cmark}
&\multicolumn{1}{p{0.5cm}}{\centering \cmark}
&\multicolumn{1}{p{0.5cm}}{\centering \cmark}
&\multicolumn{1}{p{0.5cm}}{\centering  \xmark}
&\multicolumn{1}{p{0.5cm}}{\centering -}
&\multicolumn{1}{p{0.5cm}}{\centering \xmark}
\\[6pt]\\

\multicolumn{1}{p{1cm}}{\centering Haddaji et al.\cite{haddaji2018backhauling}}
&\multicolumn{1}{p{2cm}}{\centering Proposed a cost of ownership BackHauling-as-a-Service analysis method}
&\multicolumn{1}{p{0.5cm}}{\centering \cmark}
&\multicolumn{1}{p{0.5cm}}{\centering \xmark}
&\multicolumn{1}{p{0.5cm}}{\centering  \xmark}
&\multicolumn{1}{p{0.5cm}}{\centering \cmark}
&\multicolumn{1}{p{0.5cm}}{\centering \cmark}
&\multicolumn{1}{p{0.5cm}}{\centering \xmark}
&\multicolumn{1}{p{0.5cm}}{\centering \xmark}
&\multicolumn{1}{p{0.5cm}}{\centering \xmark}
&\multicolumn{1}{p{0.5cm}}{\centering  \cmark}
&\multicolumn{1}{p{0.5cm}}{\centering \cmark}
&\multicolumn{1}{p{0.5cm}}{\centering \xmark}
&\multicolumn{1}{p{0.5cm}}{\centering -}
&\multicolumn{1}{p{0.5cm}}{\centering \xmark}
&\multicolumn{1}{p{0.5cm}}{\centering \xmark}
&\multicolumn{1}{p{0.5cm}}{\centering  \xmark}
&\multicolumn{1}{p{0.5cm}}{\centering \xmark}
&\multicolumn{1}{p{0.5cm}}{\centering \xmark}
\\[6pt]\\

\hline

\end{longtable}%}
\end{center}
\end{landscape}

\subsubsection{SDN-based Frameworks and Solutions}
Every specific requirement of 5G mobile backhaul networks emphasized the performance and cost of enabling technologies. Several significant research on SDN-based frameworks and solutions for 5G mobile backhaul have been published as presented in Table~\ref{Tablesdn}.

Perez et al. \cite{ costa20175g} focused on the 5G transport solutions and presented a new transport network architecture for the integration of fronthaul and backhaul. The study emphasized that the adoption of the new architecture with existing solutions has not increased the complexities but maximized the compatibility and supported better resource management facilities. Furthermore, the use cases of the integration of fronthaul and backhaul are additionally discussed. To provide a better flexibility for supporting different architectures, Jungnickel et al. \cite{jungnickel2014software} presented a software-defined RAN for the fronthaul and backhaul in 5G. The authors also discussed the distributed security for network slicing. The authors considered both centralized and decentralized architectures for 5G networks.

Requena et al.\cite{costa2014software} focused on the benefits of adoption of SDN-based solution with integrations of the 3GPP mobile architecture. The authors suggested SDN-based solutions for various applications like backhaul, mobility, caching, traffic, and resource management, which brought new business models and opportunities in the 5G networks. Hu et al.\cite{hu2015mih} provided a description for vertical handover framework for reducing the handover latency. The authors discussed the capacity and efficiency of their architecture as well as the power consumptions.  Furthermore, a Media-independent protocol for handover decisions to reduce latency for excessive mobility users is also discussed by the authors. Niephaus et al.\cite{niephaus2015wireless} presented an SDN-based backhaul network architecture emphasizing on the self-managing facilities and QoS. The authors emphasized the Wireless Backhaul (WiBACK) technology, which provides a smart and flexible backhaul network.

The increase in traffic over the network requires a sustainable solution for adequate traffic monitoring and load balancing to support the 5G networks. The existing prominent solutions concentrate on the core network, which calls for some flexible solutions to support backhaul and other networks while leveraging the core network services. In the era of traffic engineering, Wang et al.\cite{wang2017efficient} described the traffic engineering problem of 5G core and backhaul networks. The authors further enlightened various aspects of traffic flow with data gateways like an ideal flow of data to the gateway, multiple base stations to single data gateway and multiple flows from a single data gateway. The authors focused on the link utilization problems and multi-commodity flow problems of the network and proposed an efficient i-FPTAS algorithm for 5G mobile traffic engineering. Liyanage et al. \cite{liyanage2017software} studied the Software-Defined Monitoring (SDM) on backhaul networks to overcome the limitations of the existing monitoring systems, such as the effect of a single change in the overall system and the increased provisioning and operational costs due to the distributed infrastructure. The authors also discussed the compatibility challenges of the existing monitoring systems as well as CapEx and OpEx costs of network monitoring on the adoption of SDN architectures. Furthermore, Muñoz et al. \cite{munoz2018sdn} presented hierarchical transport SDN control and monitoring architecture emphasizing on the packet transport solutions. To reduce the administrative cost for operators and multi-domain dependencies, like scalability, modularity, and security of the networks, hierarchical Transport SDN control approach is suggested by the authors.

Requena et al.\cite{costa2015software} discussed usage of SDN in mobile backhaul for 5G networks. As stated by the authors, the adoption of SDN in future next-generation networks will enhance the backhaul network performance with a greater degree of service awareness and optimum use of network resources. Bercovich et al.\cite{bercovich2015software} discussed the wireless backhaul level transport networks. Seppanen et al.\cite{seppanen2015integrating} emphasized the integration of SDN-based transport network control and WMN backhaul. The authors suggested a mmWave wireless mesh networks backhaul solution with SDN-based centralized transport network control to avoid ambiguity among local and centralized controller and increase the resource sharing capabilities.

\begin{landscape}
\begin{center}
\fontsize{6}{6}\selectfont
\setlength\LTleft{10pt}            % default: \parindent
\setlength\LTright{0pt}
%\begin{longtable}{llllllllllll}
\begin{longtable}{@{\extracolsep{\fill}}*{19}{c}}
\caption{A comparison of existing SDN-based frameworks and solutions for 5G mobile backhaul networks.(R1:Latency/Delay Consideration, R2:Capacity Enhancement, R3:Quality of
Service, R4:Easy Installation/ Cost Effectiveness factors Considerations, R5:Traffic Management, R6:Synchronization, R7:Energy Considerations, R8:Complexity Consideration, R9:Fronthaul, R10:Multi-tenancy/Network Slicing support, R11:Handover Support, R12:Link Utilization, R13:Load Balancing, R14:RAN, R15:SON, R16:MAC Protocol Considerations, R17:Considerations of Stands )}\label{Tablesdn} \\
\hline
\\[6pt]\\
\multicolumn{1}{p{1cm}}{\centering \textbf{Scheme}}
&\multicolumn{1}{p{2cm}}{\centering \textbf{Main Contributions}}
&\multicolumn{1}{p{0.5cm}}{\centering \textbf{R1} }
&\multicolumn{1}{p{0.5cm}}{\centering \textbf{R2}}
&\multicolumn{1}{p{0.5cm}}{\centering \textbf{R3}}
&\multicolumn{1}{p{0.5cm}}{\centering \textbf{R4}}
&\multicolumn{1}{p{0.5cm}}{\centering \textbf{R5}}
&\multicolumn{1}{p{0.5cm}}{\centering \textbf{R6}}
&\multicolumn{1}{p{0.5cm}}{\centering \textbf{R7}}
&\multicolumn{1}{p{0.5cm}}{\centering \textbf{R8}}
&\multicolumn{1}{p{0.5cm}}{\centering \textbf{R9}}
&\multicolumn{1}{p{0.5cm}}{\centering \textbf{R10}}
&\multicolumn{1}{p{0.5cm}}{\centering \textbf{R11}}
&\multicolumn{1}{p{0.5cm}}{\centering \textbf{R12}}
&\multicolumn{1}{p{0.5cm}}{\centering \textbf{R13}}
&\multicolumn{1}{p{0.5cm}}{\centering \textbf{R14}}
&\multicolumn{1}{p{0.5cm}}{\centering \textbf{R15}}
&\multicolumn{1}{p{0.5cm}}{\centering \textbf{R16}}
&\multicolumn{1}{p{0.5cm}}{\centering \textbf{R17}}
\\[6pt]\\
\hline \\%data entry for gu

\endfirsthead

\multicolumn{19}{c}%
{{\bfseries \tablename\ \thetable{} -- continued from previous page}} \\
\\[6pt]\\
\hline
\\[6pt]\\
\multicolumn{1}{p{1cm}}{\centering \textbf{Scheme}}
&\multicolumn{1}{p{2cm}}{\centering \textbf{Main Contributions}}
&\multicolumn{1}{p{0.5cm}}{\centering \textbf{R1} }
&\multicolumn{1}{p{0.5cm}}{\centering \textbf{R2}}
&\multicolumn{1}{p{0.5cm}}{\centering \textbf{R3}}
&\multicolumn{1}{p{0.5cm}}{\centering \textbf{R4}}
&\multicolumn{1}{p{0.5cm}}{\centering \textbf{R5}}
&\multicolumn{1}{p{0.5cm}}{\centering \textbf{R6}}
&\multicolumn{1}{p{0.5cm}}{\centering \textbf{R7}}
&\multicolumn{1}{p{0.5cm}}{\centering \textbf{R8}}
&\multicolumn{1}{p{0.5cm}}{\centering \textbf{R9}}
&\multicolumn{1}{p{0.5cm}}{\centering \textbf{R10}}
&\multicolumn{1}{p{0.5cm}}{\centering \textbf{R11}}
&\multicolumn{1}{p{0.5cm}}{\centering \textbf{R12}}
&\multicolumn{1}{p{0.5cm}}{\centering \textbf{R13}}
&\multicolumn{1}{p{0.5cm}}{\centering \textbf{R14}}
&\multicolumn{1}{p{0.5cm}}{\centering \textbf{R15}}
&\multicolumn{1}{p{0.5cm}}{\centering \textbf{R16}}
&\multicolumn{1}{p{0.5cm}}{\centering \textbf{R17}}
\\[6pt]\\
\hline \\
\endhead

\hline \multicolumn{13}{l}{{Continued on next page}} \\

\endfoot

\endlastfoot

\multicolumn{1}{p{1cm}}{\centering Perez et al. \cite{ costa20175g}}
&\multicolumn{1}{p{2cm}}{\centering Presented a new transport network architecture for the integration of fronthaul and backhaul}
&\multicolumn{1}{p{0.5cm}}{\centering \cmark}
&\multicolumn{1}{p{0.5cm}}{\centering \xmark}
&\multicolumn{1}{p{0.5cm}}{\centering  \cmark}
&\multicolumn{1}{p{0.5cm}}{\centering \cmark}
&\multicolumn{1}{p{0.5cm}}{\centering \cmark}
&\multicolumn{1}{p{0.5cm}}{\centering \xmark}
&\multicolumn{1}{p{0.5cm}}{\centering -}
&\multicolumn{1}{p{0.5cm}}{\centering -}
&\multicolumn{1}{p{0.5cm}}{\centering  \cmark}
&\multicolumn{1}{p{0.5cm}}{\centering \cmark}
&\multicolumn{1}{p{0.5cm}}{\centering \xmark}
&\multicolumn{1}{p{0.5cm}}{\centering -}
&\multicolumn{1}{p{0.5cm}}{\centering \cmark}
&\multicolumn{1}{p{0.5cm}}{\centering \cmark}
&\multicolumn{1}{p{0.5cm}}{\centering  \xmark}
&\multicolumn{1}{p{0.5cm}}{\centering \cmark}
&\multicolumn{1}{p{0.5cm}}{\centering \cmark}
\\[6pt]\\

\multicolumn{1}{p{1cm}}{\centering Jungnickel et al. \cite{jungnickel2014software}}
&\multicolumn{1}{p{2cm}}{\centering Presented Software defined RAN for the fronthaul and backhaul in 5G}
&\multicolumn{1}{p{0.5cm}}{\centering \cmark}
&\multicolumn{1}{p{0.5cm}}{\centering \cmark}
&\multicolumn{1}{p{0.5cm}}{\centering  \xmark}
&\multicolumn{1}{p{0.5cm}}{\centering \cmark}
&\multicolumn{1}{p{0.5cm}}{\centering \cmark}
&\multicolumn{1}{p{0.5cm}}{\centering \cmark}
&\multicolumn{1}{p{0.5cm}}{\centering -}
&\multicolumn{1}{p{0.5cm}}{\centering -}
&\multicolumn{1}{p{0.5cm}}{\centering  \cmark}
&\multicolumn{1}{p{0.5cm}}{\centering \xmark}
&\multicolumn{1}{p{0.5cm}}{\centering \cmark}
&\multicolumn{1}{p{0.5cm}}{\centering -}
&\multicolumn{1}{p{0.5cm}}{\centering \cmark}
&\multicolumn{1}{p{0.5cm}}{\centering \cmark}
&\multicolumn{1}{p{0.5cm}}{\centering  \xmark}
&\multicolumn{1}{p{0.5cm}}{\centering \xmark}
&\multicolumn{1}{p{0.5cm}}{\centering -}
 \\[6pt]\\

\multicolumn{1}{p{1cm}}{\centering Requena et al.\cite{costa2014software}}
&\multicolumn{1}{p{2cm}}{\centering Discussed the benefits of SDN in 5G and migration of 3GPP mobile architecture to SDN}
&\multicolumn{1}{p{0.5cm}}{\centering \cmark}
&\multicolumn{1}{p{0.5cm}}{\centering \cmark}
&\multicolumn{1}{p{0.5cm}}{\centering  \cmark}
&\multicolumn{1}{p{0.5cm}}{\centering \cmark}
&\multicolumn{1}{p{0.5cm}}{\centering \cmark}
&\multicolumn{1}{p{0.5cm}}{\centering \xmark}
&\multicolumn{1}{p{0.5cm}}{\centering -}
&\multicolumn{1}{p{0.5cm}}{\centering -}
&\multicolumn{1}{p{0.5cm}}{\centering  \xmark}
&\multicolumn{1}{p{0.5cm}}{\centering \xmark}
&\multicolumn{1}{p{0.5cm}}{\centering \cmark}
&\multicolumn{1}{p{0.5cm}}{\centering -}
&\multicolumn{1}{p{0.5cm}}{\centering \cmark}
&\multicolumn{1}{p{0.5cm}}{\centering \xmark}
&\multicolumn{1}{p{0.5cm}}{\centering  \cmark}
&\multicolumn{1}{p{0.5cm}}{\centering \cmark}
&\multicolumn{1}{p{0.5cm}}{\centering \cmark}   \\[6pt]\\

\multicolumn{1}{p{1cm}}{\centering Hu et al.\cite{hu2015mih}}
&\multicolumn{1}{p{2cm}}{\centering Proposed a vertical
handover framework for reducing handover latency}
&\multicolumn{1}{p{0.5cm}}{\centering \cmark}
&\multicolumn{1}{p{0.5cm}}{\centering -}
&\multicolumn{1}{p{0.5cm}}{\centering  \cmark}
&\multicolumn{1}{p{0.5cm}}{\centering \xmark}
&\multicolumn{1}{p{0.5cm}}{\centering \cmark}
&\multicolumn{1}{p{0.5cm}}{\centering \xmark}
&\multicolumn{1}{p{0.5cm}}{\centering \cmark}
&\multicolumn{1}{p{0.5cm}}{\centering \cmark}
&\multicolumn{1}{p{0.5cm}}{\centering \xmark}
&\multicolumn{1}{p{0.5cm}}{\centering \xmark}
&\multicolumn{1}{p{0.5cm}}{\centering \cmark}
&\multicolumn{1}{p{0.5cm}}{\centering \xmark}
&\multicolumn{1}{p{0.5cm}}{\centering \cmark}
&\multicolumn{1}{p{0.5cm}}{\centering \cmark}
&\multicolumn{1}{p{0.5cm}}{\centering \xmark}
&\multicolumn{1}{p{0.5cm}}{\centering \cmark}
&\multicolumn{1}{p{0.5cm}}{\centering \cmark} \\[6pt]\\

\multicolumn{1}{p{1cm}}{\centering Niephaus et al.\cite{niephaus2015wireless}}
&\multicolumn{1}{p{2cm}}{\centering Proposed a SDN-based backhaul network architecture for 5G}
&\multicolumn{1}{p{0.5cm}}{\centering \cmark}
&\multicolumn{1}{p{0.5cm}}{\centering \cmark}
&\multicolumn{1}{p{0.5cm}}{\centering  \cmark}
&\multicolumn{1}{p{0.5cm}}{\centering -}
&\multicolumn{1}{p{0.5cm}}{\centering \cmark}
&\multicolumn{1}{p{0.5cm}}{\centering \xmark}
&\multicolumn{1}{p{0.5cm}}{\centering \xmark}
&\multicolumn{1}{p{0.5cm}}{\centering \xmark}
&\multicolumn{1}{p{0.5cm}}{\centering \xmark}
&\multicolumn{1}{p{0.5cm}}{\centering \xmark}
&\multicolumn{1}{p{0.5cm}}{\centering \xmark}
&\multicolumn{1}{p{0.5cm}}{\centering -}
&\multicolumn{1}{p{0.5cm}}{\centering \xmark}
&\multicolumn{1}{p{0.5cm}}{\centering \cmark}
&\multicolumn{1}{p{0.5cm}}{\centering \xmark}
&\multicolumn{1}{p{0.5cm}}{\centering \xmark}
&\multicolumn{1}{p{0.5cm}}{\centering \cmark}   \\[6pt]\\\\

\multicolumn{1}{p{1cm}}{\centering Wang et al.\cite{wang2017efficient}}
&\multicolumn{1}{p{2cm}}{\centering Emphasized on the traffic engineering problem of 5G core and backhaul network}
&\multicolumn{1}{p{0.5cm}}{\centering \xmark}
&\multicolumn{1}{p{0.5cm}}{\centering \cmark}
&\multicolumn{1}{p{0.5cm}}{\centering  \cmark}
&\multicolumn{1}{p{0.5cm}}{\centering \cmark}
&\multicolumn{1}{p{0.5cm}}{\centering \cmark}
&\multicolumn{1}{p{0.5cm}}{\centering \xmark}
&\multicolumn{1}{p{0.5cm}}{\centering -}
&\multicolumn{1}{p{0.5cm}}{\centering \cmark}
&\multicolumn{1}{p{0.5cm}}{\centering \xmark}
&\multicolumn{1}{p{0.5cm}}{\centering \xmark}
&\multicolumn{1}{p{0.5cm}}{\centering \cmark}
&\multicolumn{1}{p{0.5cm}}{\centering \cmark}
&\multicolumn{1}{p{0.5cm}}{\centering -}
&\multicolumn{1}{p{0.5cm}}{\centering \cmark}
&\multicolumn{1}{p{0.5cm}}{\centering \xmark}
&\multicolumn{1}{p{0.5cm}}{\centering \xmark}
&\multicolumn{1}{p{0.5cm}}{\centering -} \\[6pt]\\

\multicolumn{1}{p{1cm}}{\centering Liyanage et al. \cite{liyanage2017software}}
&\multicolumn{1}{p{2cm}}{\centering Focused on the software defined monitoring on backhaul networks}
&\multicolumn{1}{p{0.5cm}}{\centering \cmark}
&\multicolumn{1}{p{0.5cm}}{\centering \xmark}
&\multicolumn{1}{p{0.5cm}}{\centering  -}
&\multicolumn{1}{p{0.5cm}}{\centering \cmark}
&\multicolumn{1}{p{0.5cm}}{\centering \cmark}
&\multicolumn{1}{p{0.5cm}}{\centering \xmark}
&\multicolumn{1}{p{0.5cm}}{\centering \xmark}
&\multicolumn{1}{p{0.5cm}}{\centering \xmark}
&\multicolumn{1}{p{0.5cm}}{\centering \xmark}
&\multicolumn{1}{p{0.5cm}}{\centering \xmark}
&\multicolumn{1}{p{0.5cm}}{\centering \xmark}
&\multicolumn{1}{p{0.5cm}}{\centering \xmark}
&\multicolumn{1}{p{0.5cm}}{\centering \cmark}
&\multicolumn{1}{p{0.5cm}}{\centering \xmark}
&\multicolumn{1}{p{0.5cm}}{\centering \xmark}
&\multicolumn{1}{p{0.5cm}}{\centering \xmark}
&\multicolumn{1}{p{0.5cm}}{\centering \cmark} \\[6pt]\\

\multicolumn{1}{p{1cm}}{\centering Muñoz et al. \cite{munoz2018sdn}}
&\multicolumn{1}{p{2cm}}{\centering Presented hierarchical transport SDN control and monitoring architecture emphasizing on the packet transport solutions.}
&\multicolumn{1}{p{0.5cm}}{\centering \xmark}
&\multicolumn{1}{p{0.5cm}}{\centering \cmark}
&\multicolumn{1}{p{0.5cm}}{\centering  \xmark}
&\multicolumn{1}{p{0.5cm}}{\centering \cmark}
&\multicolumn{1}{p{0.5cm}}{\centering \cmark}
&\multicolumn{1}{p{0.5cm}}{\centering \xmark}
&\multicolumn{1}{p{0.5cm}}{\centering \xmark}
&\multicolumn{1}{p{0.5cm}}{\centering -}
&\multicolumn{1}{p{0.5cm}}{\centering \cmark}
&\multicolumn{1}{p{0.5cm}}{\centering \cmark}
&\multicolumn{1}{p{0.5cm}}{\centering \cmark}
&\multicolumn{1}{p{0.5cm}}{\centering \xmark}
&\multicolumn{1}{p{0.5cm}}{\centering \xmark}
&\multicolumn{1}{p{0.5cm}}{\centering \xmark}
&\multicolumn{1}{p{0.5cm}}{\centering \xmark}
&\multicolumn{1}{p{0.5cm}}{\centering \xmark}
&\multicolumn{1}{p{0.5cm}}{\centering \cmark}  \\[6pt]\\

\multicolumn{1}{p{1cm}}{\centering Requena et al.\cite{costa2015software}}
&\multicolumn{1}{p{2cm}}{\centering Discussed usage of SDN in mobile backhaul for 5G networks}
&\multicolumn{1}{p{0.5cm}}{\centering \cmark}
&\multicolumn{1}{p{0.5cm}}{\centering \xmark}
&\multicolumn{1}{p{0.5cm}}{\centering  \cmark}
&\multicolumn{1}{p{0.5cm}}{\centering \xmark}
&\multicolumn{1}{p{0.5cm}}{\centering \cmark}
&\multicolumn{1}{p{0.5cm}}{\centering \xmark}
&\multicolumn{1}{p{0.5cm}}{\centering \cmark}
&\multicolumn{1}{p{0.5cm}}{\centering \xmark}
&\multicolumn{1}{p{0.5cm}}{\centering \xmark}
&\multicolumn{1}{p{0.5cm}}{\centering \cmark}
&\multicolumn{1}{p{0.5cm}}{\centering \cmark}
&\multicolumn{1}{p{0.5cm}}{\centering \xmark}
&\multicolumn{1}{p{0.5cm}}{\centering \xmark}
&\multicolumn{1}{p{0.5cm}}{\centering \xmark}
&\multicolumn{1}{p{0.5cm}}{\centering \xmark}
&\multicolumn{1}{p{0.5cm}}{\centering \xmark}
&\multicolumn{1}{p{0.5cm}}{\centering -}   \\[6pt]\\

\multicolumn{1}{p{1cm}}{\centering Bercovich et al.\cite{bercovich2015software}}
&\multicolumn{1}{p{2cm}}{\centering Discussed the wireless backhaul level transport networks}
&\multicolumn{1}{p{0.5cm}}{\centering \cmark}
&\multicolumn{1}{p{0.5cm}}{\centering \cmark}
&\multicolumn{1}{p{0.5cm}}{\centering  -}
&\multicolumn{1}{p{0.5cm}}{\centering \cmark}
&\multicolumn{1}{p{0.5cm}}{\centering \cmark}
&\multicolumn{1}{p{0.5cm}}{\centering \xmark}
&\multicolumn{1}{p{0.5cm}}{\centering \cmark}
&\multicolumn{1}{p{0.5cm}}{\centering -}
&\multicolumn{1}{p{0.5cm}}{\centering \cmark}
&\multicolumn{1}{p{0.5cm}}{\centering \xmark}
&\multicolumn{1}{p{0.5cm}}{\centering \xmark}
&\multicolumn{1}{p{0.5cm}}{\centering \cmark}
&\multicolumn{1}{p{0.5cm}}{\centering \xmark}
&\multicolumn{1}{p{0.5cm}}{\centering -}
&\multicolumn{1}{p{0.5cm}}{\centering \xmark}
&\multicolumn{1}{p{0.5cm}}{\centering \xmark}
&\multicolumn{1}{p{0.5cm}}{\centering \cmark}
   \\[6pt]\\

\multicolumn{1}{p{1cm}}{\centering Seppanen et al.\cite{seppanen2015integrating}}
&\multicolumn{1}{p{2cm}}{\centering Integrate the SDN-based transport network control and WMN backhaul}
&\multicolumn{1}{p{0.5cm}}{\centering \cmark}
&\multicolumn{1}{p{0.5cm}}{\centering \cmark}
&\multicolumn{1}{p{0.5cm}}{\centering  \xmark}
&\multicolumn{1}{p{0.5cm}}{\centering -}
&\multicolumn{1}{p{0.5cm}}{\centering \cmark}
&\multicolumn{1}{p{0.5cm}}{\centering \xmark}
&\multicolumn{1}{p{0.5cm}}{\centering \xmark}
&\multicolumn{1}{p{0.5cm}}{\centering \xmark}
&\multicolumn{1}{p{0.5cm}}{\centering \xmark}
&\multicolumn{1}{p{0.5cm}}{\centering \cmark}
&\multicolumn{1}{p{0.5cm}}{\centering -}
&\multicolumn{1}{p{0.5cm}}{\centering \cmark}
&\multicolumn{1}{p{0.5cm}}{\centering \cmark}
&\multicolumn{1}{p{0.5cm}}{\centering \xmark}
&\multicolumn{1}{p{0.5cm}}{\centering \xmark}
&\multicolumn{1}{p{0.5cm}}{\centering \cmark}
&\multicolumn{1}{p{0.5cm}}{\centering -} \\[6pt]\\

\hline

\end{longtable}%}
\end{center}
\end{landscape}

\subsubsection{mmWave-Based Frameworks and Solutions}
The mmWave communication is an emerging technology to support a very dense deployment of mmWave base stations in the 5G networks. The mmWave provides flexibilities and reduces cost in the 5G mobile backhaul as well as improves the spectral efficiency of the access link through highly directional beamforming.  The integration of large bandwidth and advanced RF beam forming that supports mmWave cellular system are prominent backhaul solutions in 5G networks. From the existing literature, the mmWave-based backhaul solution represents itself a strong candidate for the 5G networks. Mesodiakaki et al.\cite{ mesodiakaki2018optimal} defined a policy-based framework for energy optimization for user association, routing, and backhaul switching. The authors demonstrated an algorithm to calculate the most energy-efficient User Equipment (UE) association and backhaul routing strategy. The authors further discussed the feasibility of the proposed solution with SDN-based backhaul solutions. Saadat et al. \cite{ saadat2018multipath} described a distributed routing scheme for backhaul traffic. The authors presented a comparison between multiple-association, multihop, and multi path-multi hop backhaul schemes in the ultra-dense networks. Gao et al.\cite{ gao2015mmwave} focused on the ultra-dense networks and mmWave challenges and advantages. The authors highlighted the advantages such as band utilization and easy deployment of the antenna for the mmWave.

The efficient routing and scheduling paradigms boost the 5G backhaul network performance. Yuan et al.\cite{ yuan2018optimal} focused on the routing and scheduling to improve backhaul efficiency. The scheduling is done based on the matching theory, and efficient edge-coloring based approximation algorithm is proposed to enhance the throughput. Saha et al.\cite{ saha2018bandwidth} emphasized on the Bandwidth strategies and gives a tractable model integrated access and backhaul enabled mmWave networks. The authors consider three bandwidth partition schemes, namely, equal partitioning, instant partitioning, and average partitioning scheme. Taori and  Sridharan\cite{ taori2014band} described point to multi-point architecture with the related cost and latency issues. The authors discussed the TDM-based scheduling schemes for the backhauls and argued about the feasibility of mmWave at a tolerable loss in the in-band solutions. The concepts of mmWave links for backhaul and associated challenges such as resource management, beam forming and beam tracking were discussed by the Weiler et al.\cite{weiler2014enabling}. The authors emphasized the control/user plane split in the 5G for the high data rates.

Singh et al.\cite{ singh2015tractable} presented a tractable model for rate distribution in mmWave cellular networks. The self-backhauling concept based on the saturation density was also emphasized. Furthermore, Pham et al.\cite{pham2015hybrid} described the integration of mmWave and FSO links with 5G backhaul architecture. Jung and Lee\cite{jung2016outage} focused on the outage probability of a wireless backhaul. Li et al.\cite{li2017radio} emphasized the joint backhaul and radio resource management. The authors also provided joint scheduling and resource allocation algorithm by considering the computational complexity and resource utilization. Furthermore, the authors discussed the challenges and requirements of the backhauling in mmWave RAN. They also considered the self-backhauling concepts of wireless backhaul as a potential solution in the coverage and capacity enhancement. Nasr and Fahmy\cite{nasr2017millimeter} presented the comparisons of mesh and star topologies of mmWave wireless backhauling for 5G through scalability, reliability and fairness factors. The authors presented the effects on the throughput by varying cell size and the number of cells. Destino et al.\cite{destino2017system} considered the RF component requirements for mmWave mobile backhaul transceivers. Hu and Blough\cite{hu2017relay} offered a new mmWave backhaul network architecture and algorithm for relay selection and scheduling to enhance the end-to-end throughput.

The economic consideration of mmWave to cope up with 5G backhaul is also a significant challenge. The operator side and deployment side considered various feasibility issues on the adoption of mmWave, among them, economic related is one of the major issues. Magne et al.\cite{magne2018millimeter} focused on the cost of ownership and flexibility of the mmWave point to multipoint in backhaul. The authors presented cost wise comparisons of W-band Point-to-multipoint (PtmP) to E-band PtmP.

\begin{landscape}
\begin{center}
\fontsize{6}{6}\selectfont
\setlength\LTleft{10pt}            % default: \parindent
\setlength\LTright{0pt}
%\begin{longtable}{llllllllllll}
\begin{longtable}{@{\extracolsep{\fill}}*{23}{c}}
\caption{A comparison of existing mmWave-based frameworks and solutions for 5G Mobile backhaul networks.(R1:Latency/Delay Consideration, R2:Capacity Enhancement, R3:Quality of
Service, R4:Easy Installation/Cost Effectiveness factors Considerations, R5:Traffic Management, R6:Synchronization, R7:Energy Considerations, R8:Complexity Considerations, R9:Fronthaul, R10:Multi-tenancy/Network Slicing support, R11:Handover Supports, R12:Link Utilization, R13:Load Balancing, R14:RAN, R15:SON, R16:MAC Protocol Considerations, R17:Consideration of Standards, R18:Path Loss
Considerations, R19:Scheduling, R20:Point-to-Point, R21:Point-to-Multi point)}\label{Tablemmwave} \\
\hline
\\[6pt]\\
\multicolumn{1}{p{1cm}}{\centering \textbf{Scheme}}
&\multicolumn{1}{p{2cm}}{\centering \textbf{Main Contributions}}
&\multicolumn{1}{p{0.5cm}}{\centering \textbf{R1} }
&\multicolumn{1}{p{0.5cm}}{\centering \textbf{R2}}
&\multicolumn{1}{p{0.5cm}}{\centering \textbf{R3}}
&\multicolumn{1}{p{0.5cm}}{\centering \textbf{R4}}
&\multicolumn{1}{p{0.5cm}}{\centering \textbf{R5}}
&\multicolumn{1}{p{0.5cm}}{\centering \textbf{R6}}
&\multicolumn{1}{p{0.5cm}}{\centering \textbf{R7}}
&\multicolumn{1}{p{0.5cm}}{\centering \textbf{R8}}
&\multicolumn{1}{p{0.5cm}}{\centering \textbf{R9}}
&\multicolumn{1}{p{0.5cm}}{\centering \textbf{R10}}
&\multicolumn{1}{p{0.5cm}}{\centering \textbf{R11}}
&\multicolumn{1}{p{0.5cm}}{\centering \textbf{R12}}
&\multicolumn{1}{p{0.5cm}}{\centering \textbf{R13}}
&\multicolumn{1}{p{0.5cm}}{\centering \textbf{R14}}
&\multicolumn{1}{p{0.5cm}}{\centering \textbf{R15}}
&\multicolumn{1}{p{0.5cm}}{\centering \textbf{R16}}
&\multicolumn{1}{p{0.5cm}}{\centering \textbf{R17}}
&\multicolumn{1}{p{0.5cm}}{\centering \textbf{R18}}
&\multicolumn{1}{p{0.5cm}}{\centering \textbf{R19}}
&\multicolumn{1}{p{0.5cm}}{\centering \textbf{R20}}
&\multicolumn{1}{p{0.5cm}}{\centering \textbf{R21}}
\\[6pt]\\
\hline \\%data entry for gu

\endfirsthead

\multicolumn{23}{c}%
{{\bfseries \tablename\ \thetable{} -- continued from previous page}} \\
\hline
\\[6pt]\\
\multicolumn{1}{p{1cm}}{\centering \textbf{Scheme}}
&\multicolumn{1}{p{2cm}}{\centering \textbf{Main Contributions}}
&\multicolumn{1}{p{0.5cm}}{\centering \textbf{R1} }
&\multicolumn{1}{p{0.5cm}}{\centering \textbf{R2}}
&\multicolumn{1}{p{0.5cm}}{\centering \textbf{R3}}
&\multicolumn{1}{p{0.5cm}}{\centering \textbf{R4}}
&\multicolumn{1}{p{0.5cm}}{\centering \textbf{R5}}
&\multicolumn{1}{p{0.5cm}}{\centering \textbf{R6}}
&\multicolumn{1}{p{0.5cm}}{\centering \textbf{R7}}
&\multicolumn{1}{p{0.5cm}}{\centering \textbf{R8}}
&\multicolumn{1}{p{0.5cm}}{\centering \textbf{R9}}
&\multicolumn{1}{p{0.5cm}}{\centering \textbf{R10}}
&\multicolumn{1}{p{0.5cm}}{\centering \textbf{R11}}
&\multicolumn{1}{p{0.5cm}}{\centering \textbf{R12}}
&\multicolumn{1}{p{0.5cm}}{\centering \textbf{R13}}
&\multicolumn{1}{p{0.5cm}}{\centering \textbf{R14}}
&\multicolumn{1}{p{0.5cm}}{\centering \textbf{R15}}
&\multicolumn{1}{p{0.5cm}}{\centering \textbf{R16}}
&\multicolumn{1}{p{0.5cm}}{\centering \textbf{R17}}
&\multicolumn{1}{p{0.5cm}}{\centering \textbf{R18}}
&\multicolumn{1}{p{0.5cm}}{\centering \textbf{R19}}
&\multicolumn{1}{p{0.5cm}}{\centering \textbf{R20}}
&\multicolumn{1}{p{0.5cm}}{\centering \textbf{R21}}
\\[6pt]\\
\hline \\
\endhead

\hline \multicolumn{23}{l}{{Continued on next page}} \\

\endfoot

\endlastfoot

\multicolumn{1}{p{1cm}}{\centering Mesodiakaki et al.\cite{ mesodiakaki2018optimal}}
&\multicolumn{1}{p{2cm}}{\centering Defined a policy-based framework for energy optimization for user association, routing and backhaul switching }
&\multicolumn{1}{p{0.5cm}}{\centering \xmark}
&\multicolumn{1}{p{0.5cm}}{\centering \cmark}
&\multicolumn{1}{p{0.5cm}}{\centering  \cmark}
&\multicolumn{1}{p{0.5cm}}{\centering \cmark}
&\multicolumn{1}{p{0.5cm}}{\centering \cmark}
&\multicolumn{1}{p{0.5cm}}{\centering \xmark}
&\multicolumn{1}{p{0.5cm}}{\centering \cmark}
&\multicolumn{1}{p{0.5cm}}{\centering \cmark}
&\multicolumn{1}{p{0.5cm}}{\centering  \xmark}
&\multicolumn{1}{p{0.5cm}}{\centering \xmark}
&\multicolumn{1}{p{0.5cm}}{\centering \xmark}
&\multicolumn{1}{p{0.5cm}}{\centering \cmark}
&\multicolumn{1}{p{0.5cm}}{\centering \cmark}
&\multicolumn{1}{p{0.5cm}}{\centering \xmark}
&\multicolumn{1}{p{0.5cm}}{\centering  -}
&\multicolumn{1}{p{0.5cm}}{\centering \xmark}
&\multicolumn{1}{p{0.5cm}}{\centering \xmark}
&\multicolumn{1}{p{0.5cm}}{\centering \cmark}
&\multicolumn{1}{p{0.5cm}}{\centering  \xmark}
&\multicolumn{1}{p{0.5cm}}{\centering \xmark}
&\multicolumn{1}{p{0.5cm}}{\centering \xmark}
\\[6pt]\\

\multicolumn{1}{p{1cm}}{\centering Saadat et al. \cite{ saadat2018multipath}}
&\multicolumn{1}{p{2cm}}{\centering Proposed a distributed routing scheme for backhaul traffic}
&\multicolumn{1}{p{0.5cm}}{\centering \cmark}
&\multicolumn{1}{p{0.5cm}}{\centering \cmark}
&\multicolumn{1}{p{0.5cm}}{\centering  \cmark}
&\multicolumn{1}{p{0.5cm}}{\centering \cmark}
&\multicolumn{1}{p{0.5cm}}{\centering \cmark}
&\multicolumn{1}{p{0.5cm}}{\centering \xmark}
&\multicolumn{1}{p{0.5cm}}{\centering \cmark}
&\multicolumn{1}{p{0.5cm}}{\centering \xmark}
&\multicolumn{1}{p{0.5cm}}{\centering  \xmark}
&\multicolumn{1}{p{0.5cm}}{\centering \xmark}
&\multicolumn{1}{p{0.5cm}}{\centering \xmark}
&\multicolumn{1}{p{0.5cm}}{\centering \cmark}
&\multicolumn{1}{p{0.5cm}}{\centering \cmark}
&\multicolumn{1}{p{0.5cm}}{\centering \xmark}
&\multicolumn{1}{p{0.5cm}}{\centering  \xmark}
&\multicolumn{1}{p{0.5cm}}{\centering \xmark}
&\multicolumn{1}{p{0.5cm}}{\centering \xmark}
&\multicolumn{1}{p{0.5cm}}{\centering \cmark}
&\multicolumn{1}{p{0.5cm}}{\centering  -}
&\multicolumn{1}{p{0.5cm}}{\centering \xmark}
&\multicolumn{1}{p{0.5cm}}{\centering \xmark}
\\[6pt]\\

\multicolumn{1}{p{1cm}}{\centering Gao et al.\cite{gao2015mmwave}}
&\multicolumn{1}{p{2cm}}{\centering Focused on the ultra-dense networks and mmWave challenges and advantages}
&\multicolumn{1}{p{0.5cm}}{\centering \cmark}
&\multicolumn{1}{p{0.5cm}}{\centering \cmark}
&\multicolumn{1}{p{0.5cm}}{\centering  \xmark}
&\multicolumn{1}{p{0.5cm}}{\centering \cmark}
&\multicolumn{1}{p{0.5cm}}{\centering \cmark}
&\multicolumn{1}{p{0.5cm}}{\centering \cmark}
&\multicolumn{1}{p{0.5cm}}{\centering \cmark}
&\multicolumn{1}{p{0.5cm}}{\centering \cmark}
&\multicolumn{1}{p{0.5cm}}{\centering  \xmark}
&\multicolumn{1}{p{0.5cm}}{\centering \xmark}
&\multicolumn{1}{p{0.5cm}}{\centering \xmark}
&\multicolumn{1}{p{0.5cm}}{\centering \cmark}
&\multicolumn{1}{p{0.5cm}}{\centering -}
&\multicolumn{1}{p{0.5cm}}{\centering \cmark}
&\multicolumn{1}{p{0.5cm}}{\centering  \xmark}
&\multicolumn{1}{p{0.5cm}}{\centering \xmark}
&\multicolumn{1}{p{0.5cm}}{\centering \xmark}
&\multicolumn{1}{p{0.5cm}}{\centering \cmark}
&\multicolumn{1}{p{0.5cm}}{\centering  \cmark}
&\multicolumn{1}{p{0.5cm}}{\centering \cmark}
&\multicolumn{1}{p{0.5cm}}{\centering \cmark}
\\[6pt]\\

\multicolumn{1}{p{1cm}}{\centering Yuan et al.\cite{ yuan2018optimal}}
&\multicolumn{1}{p{2cm}}{\centering Focused on the routing and scheduling to improve backhaul efficiency.}
&\multicolumn{1}{p{0.5cm}}{\centering \xmark}
&\multicolumn{1}{p{0.5cm}}{\centering \cmark}
&\multicolumn{1}{p{0.5cm}}{\centering  \cmark}
&\multicolumn{1}{p{0.5cm}}{\centering \cmark}
&\multicolumn{1}{p{0.5cm}}{\centering \cmark}
&\multicolumn{1}{p{0.5cm}}{\centering \xmark}
&\multicolumn{1}{p{0.5cm}}{\centering \cmark}
&\multicolumn{1}{p{0.5cm}}{\centering \cmark}
&\multicolumn{1}{p{0.5cm}}{\centering  \xmark}
&\multicolumn{1}{p{0.5cm}}{\centering \xmark}
&\multicolumn{1}{p{0.5cm}}{\centering \xmark}
&\multicolumn{1}{p{0.5cm}}{\centering \cmark}
&\multicolumn{1}{p{0.5cm}}{\centering \xmark}
&\multicolumn{1}{p{0.5cm}}{\centering \xmark}
&\multicolumn{1}{p{0.5cm}}{\centering  \xmark}
&\multicolumn{1}{p{0.5cm}}{\centering \xmark}
&\multicolumn{1}{p{0.5cm}}{\centering \xmark}
&\multicolumn{1}{p{0.5cm}}{\centering \cmark}
&\multicolumn{1}{p{0.5cm}}{\centering  \cmark}
&\multicolumn{1}{p{0.5cm}}{\centering \xmark}
&\multicolumn{1}{p{0.5cm}}{\centering \xmark}
\\[6pt]\\

\multicolumn{1}{p{1cm}}{\centering Saha et al.\cite{saha2018bandwidth}}
&\multicolumn{1}{p{2cm}}{\centering Proposed a tractable model integrated access and backhaul enabled mmWave Networks and emphasized on the Bandwidth strategies}
&\multicolumn{1}{p{0.5cm}}{\centering \xmark}
&\multicolumn{1}{p{0.5cm}}{\centering \xmark}
&\multicolumn{1}{p{0.5cm}}{\centering  \xmark}
&\multicolumn{1}{p{0.5cm}}{\centering \xmark}
&\multicolumn{1}{p{0.5cm}}{\centering \xmark}
&\multicolumn{1}{p{0.5cm}}{\centering \xmark}
&\multicolumn{1}{p{0.5cm}}{\centering -}
&\multicolumn{1}{p{0.5cm}}{\centering \cmark}
&\multicolumn{1}{p{0.5cm}}{\centering  \xmark}
&\multicolumn{1}{p{0.5cm}}{\centering \xmark}
&\multicolumn{1}{p{0.5cm}}{\centering \xmark}
&\multicolumn{1}{p{0.5cm}}{\centering \cmark}
&\multicolumn{1}{p{0.5cm}}{\centering \cmark}
&\multicolumn{1}{p{0.5cm}}{\centering \xmark}
&\multicolumn{1}{p{0.5cm}}{\centering  \xmark}
&\multicolumn{1}{p{0.5cm}}{\centering \cmark}
&\multicolumn{1}{p{0.5cm}}{\centering \cmark}
&\multicolumn{1}{p{0.5cm}}{\centering \cmark}
&\multicolumn{1}{p{0.5cm}}{\centering  \cmark}
&\multicolumn{1}{p{0.5cm}}{\centering \xmark}
&\multicolumn{1}{p{0.5cm}}{\centering \xmark}
\\[6pt]\\

\multicolumn{1}{p{1cm}}{\centering Taori and  Sridharan\cite{ taori2014band}}
&\multicolumn{1}{p{2cm}}{\centering Emphasized on Point to Multi-point architecture and considered cost and latency issues}
&\multicolumn{1}{p{0.5cm}}{\centering \cmark}
&\multicolumn{1}{p{0.5cm}}{\centering \cmark}
&\multicolumn{1}{p{0.5cm}}{\centering  \xmark}
&\multicolumn{1}{p{0.5cm}}{\centering \cmark}
&\multicolumn{1}{p{0.5cm}}{\centering \cmark}
&\multicolumn{1}{p{0.5cm}}{\centering \xmark}
&\multicolumn{1}{p{0.5cm}}{\centering \xmark}
&\multicolumn{1}{p{0.5cm}}{\centering \xmark}
&\multicolumn{1}{p{0.5cm}}{\centering  \xmark}
&\multicolumn{1}{p{0.5cm}}{\centering \xmark}
&\multicolumn{1}{p{0.5cm}}{\centering \cmark}
&\multicolumn{1}{p{0.5cm}}{\centering \cmark}
&\multicolumn{1}{p{0.5cm}}{\centering \xmark}
&\multicolumn{1}{p{0.5cm}}{\centering \xmark}
&\multicolumn{1}{p{0.5cm}}{\centering  \xmark}
&\multicolumn{1}{p{0.5cm}}{\centering \xmark}
&\multicolumn{1}{p{0.5cm}}{\centering \xmark}
&\multicolumn{1}{p{0.5cm}}{\centering \cmark}
&\multicolumn{1}{p{0.5cm}}{\centering  \cmark}
&\multicolumn{1}{p{0.5cm}}{\centering \xmark}
&\multicolumn{1}{p{0.5cm}}{\centering \cmark}
\\[6pt]\\

\multicolumn{1}{p{1cm}}{\centering Weiler et al.\cite{ weiler2014enabling}}
&\multicolumn{1}{p{2cm}}{\centering The  concepts of mmWave links for backhaul  and associated challenges like resource management and beam forming and beam tracking were discussed}
&\multicolumn{1}{p{0.5cm}}{\centering \cmark}
&\multicolumn{1}{p{0.5cm}}{\centering \cmark}
&\multicolumn{1}{p{0.5cm}}{\centering  \cmark}
&\multicolumn{1}{p{0.5cm}}{\centering \cmark}
&\multicolumn{1}{p{0.5cm}}{\centering \cmark}
&\multicolumn{1}{p{0.5cm}}{\centering \cmark}
&\multicolumn{1}{p{0.5cm}}{\centering \cmark}
&\multicolumn{1}{p{0.5cm}}{\centering -}
&\multicolumn{1}{p{0.5cm}}{\centering  \cmark}
&\multicolumn{1}{p{0.5cm}}{\centering \xmark}
&\multicolumn{1}{p{0.5cm}}{\centering \cmark}
&\multicolumn{1}{p{0.5cm}}{\centering \cmark}
&\multicolumn{1}{p{0.5cm}}{\centering \cmark}
&\multicolumn{1}{p{0.5cm}}{\centering \cmark}
&\multicolumn{1}{p{0.5cm}}{\centering  \xmark}
&\multicolumn{1}{p{0.5cm}}{\centering \cmark}
&\multicolumn{1}{p{0.5cm}}{\centering \cmark}
&\multicolumn{1}{p{0.5cm}}{\centering \cmark}
&\multicolumn{1}{p{0.5cm}}{\centering  -}
&\multicolumn{1}{p{0.5cm}}{\centering \xmark}
&\multicolumn{1}{p{0.5cm}}{\centering \xmark}
\\[6pt]\\

\multicolumn{1}{p{1cm}}{\centering Singh et al.\cite{ singh2015tractable}}
&\multicolumn{1}{p{2cm}}{\centering Proposed a tractable mmWave cellular Model}
&\multicolumn{1}{p{0.5cm}}{\centering \xmark}
&\multicolumn{1}{p{0.5cm}}{\centering \cmark}
&\multicolumn{1}{p{0.5cm}}{\centering  \cmark}
&\multicolumn{1}{p{0.5cm}}{\centering \xmark}
&\multicolumn{1}{p{0.5cm}}{\centering \xmark}
&\multicolumn{1}{p{0.5cm}}{\centering \xmark}
&\multicolumn{1}{p{0.5cm}}{\centering \xmark}
&\multicolumn{1}{p{0.5cm}}{\centering \xmark}
&\multicolumn{1}{p{0.5cm}}{\centering  \xmark}
&\multicolumn{1}{p{0.5cm}}{\centering \xmark}
&\multicolumn{1}{p{0.5cm}}{\centering \xmark}
&\multicolumn{1}{p{0.5cm}}{\centering \cmark}
&\multicolumn{1}{p{0.5cm}}{\centering \cmark}
&\multicolumn{1}{p{0.5cm}}{\centering \xmark}
&\multicolumn{1}{p{0.5cm}}{\centering  \cmark}
&\multicolumn{1}{p{0.5cm}}{\centering \xmark}
&\multicolumn{1}{p{0.5cm}}{\centering \xmark}
&\multicolumn{1}{p{0.5cm}}{\centering \cmark}
&\multicolumn{1}{p{0.5cm}}{\centering  \xmark}
&\multicolumn{1}{p{0.5cm}}{\centering \xmark}
&\multicolumn{1}{p{0.5cm}}{\centering \xmark}
\\[6pt]\\

\multicolumn{1}{p{1cm}}{\centering Pham et al.\cite{pham2015hybrid}}
&\multicolumn{1}{p{2cm}}{\centering Focused on the integration of MMW and FSO links with 5G backhaul architecture}
&\multicolumn{1}{p{0.5cm}}{\centering \xmark}
&\multicolumn{1}{p{0.5cm}}{\centering \cmark}
&\multicolumn{1}{p{0.5cm}}{\centering  \xmark}
&\multicolumn{1}{p{0.5cm}}{\centering \cmark}
&\multicolumn{1}{p{0.5cm}}{\centering -}
&\multicolumn{1}{p{0.5cm}}{\centering \xmark}
&\multicolumn{1}{p{0.5cm}}{\centering \cmark}
&\multicolumn{1}{p{0.5cm}}{\centering \xmark}
&\multicolumn{1}{p{0.5cm}}{\centering  \xmark}
&\multicolumn{1}{p{0.5cm}}{\centering \xmark}
&\multicolumn{1}{p{0.5cm}}{\centering \xmark}
&\multicolumn{1}{p{0.5cm}}{\centering \cmark}
&\multicolumn{1}{p{0.5cm}}{\centering \xmark}
&\multicolumn{1}{p{0.5cm}}{\centering \cmark}
&\multicolumn{1}{p{0.5cm}}{\centering  \xmark}
&\multicolumn{1}{p{0.5cm}}{\centering \xmark}
&\multicolumn{1}{p{0.5cm}}{\centering \xmark}
&\multicolumn{1}{p{0.5cm}}{\centering \xmark}
&\multicolumn{1}{p{0.5cm}}{\centering  \xmark}
&\multicolumn{1}{p{0.5cm}}{\centering \xmark}
&\multicolumn{1}{p{0.5cm}}{\centering \xmark}
\\[6pt]\\

\multicolumn{1}{p{1cm}}{\centering Cvetkovski et al.\cite{ cvetkovski2016next}}
&\multicolumn{1}{p{2cm}}{\centering Focused on the challenges and requirements of  backhaul}
&\multicolumn{1}{p{0.5cm}}{\centering \cmark}
&\multicolumn{1}{p{0.5cm}}{\centering \cmark}
&\multicolumn{1}{p{0.5cm}}{\centering  \cmark}
&\multicolumn{1}{p{0.5cm}}{\centering \cmark}
&\multicolumn{1}{p{0.5cm}}{\centering -}
&\multicolumn{1}{p{0.5cm}}{\centering \cmark}
&\multicolumn{1}{p{0.5cm}}{\centering \cmark}
&\multicolumn{1}{p{0.5cm}}{\centering \cmark}
&\multicolumn{1}{p{0.5cm}}{\centering  \xmark}
&\multicolumn{1}{p{0.5cm}}{\centering \xmark}
&\multicolumn{1}{p{0.5cm}}{\centering \xmark}
&\multicolumn{1}{p{0.5cm}}{\centering \cmark}
&\multicolumn{1}{p{0.5cm}}{\centering \xmark}
&\multicolumn{1}{p{0.5cm}}{\centering \xmark}
&\multicolumn{1}{p{0.5cm}}{\centering  \xmark}
&\multicolumn{1}{p{0.5cm}}{\centering \xmark}
&\multicolumn{1}{p{0.5cm}}{\centering \cmark}
&\multicolumn{1}{p{0.5cm}}{\centering \cmark}
&\multicolumn{1}{p{0.5cm}}{\centering  \xmark}
&\multicolumn{1}{p{0.5cm}}{\centering \xmark}
&\multicolumn{1}{p{0.5cm}}{\centering \xmark}
\\[6pt]\\

\multicolumn{1}{p{1cm}}{\centering Jung and Lee\cite{jung2016outage}}
&\multicolumn{1}{p{2cm}}{\centering Focused on the outage probability of a wireless backhaul}
&\multicolumn{1}{p{0.5cm}}{\centering \xmark}
&\multicolumn{1}{p{0.5cm}}{\centering -}
&\multicolumn{1}{p{0.5cm}}{\centering  \xmark}
&\multicolumn{1}{p{0.5cm}}{\centering \xmark}
&\multicolumn{1}{p{0.5cm}}{\centering \xmark}
&\multicolumn{1}{p{0.5cm}}{\centering \xmark}
&\multicolumn{1}{p{0.5cm}}{\centering \xmark}
&\multicolumn{1}{p{0.5cm}}{\centering \xmark}
&\multicolumn{1}{p{0.5cm}}{\centering  \xmark}
&\multicolumn{1}{p{0.5cm}}{\centering \xmark}
&\multicolumn{1}{p{0.5cm}}{\centering \xmark}
&\multicolumn{1}{p{0.5cm}}{\centering \cmark}
&\multicolumn{1}{p{0.5cm}}{\centering \xmark}
&\multicolumn{1}{p{0.5cm}}{\centering \xmark}
&\multicolumn{1}{p{0.5cm}}{\centering  \xmark}
&\multicolumn{1}{p{0.5cm}}{\centering \xmark}
&\multicolumn{1}{p{0.5cm}}{\centering \xmark}
&\multicolumn{1}{p{0.5cm}}{\centering \cmark}
&\multicolumn{1}{p{0.5cm}}{\centering  \xmark}
&\multicolumn{1}{p{0.5cm}}{\centering \xmark}
&\multicolumn{1}{p{0.5cm}}{\centering \xmark}
\\[6pt]\\

\multicolumn{1}{p{1cm}}{\centering Li et al.\cite{li2017radio}}
&\multicolumn{1}{p{2cm}}{\centering Emphasized on the joint backhaul and radio resource management}
&\multicolumn{1}{p{0.5cm}}{\centering \cmark}
&\multicolumn{1}{p{0.5cm}}{\centering \cmark}
&\multicolumn{1}{p{0.5cm}}{\centering  \xmark}
&\multicolumn{1}{p{0.5cm}}{\centering \xmark}
&\multicolumn{1}{p{0.5cm}}{\centering \xmark}
&\multicolumn{1}{p{0.5cm}}{\centering \xmark}
&\multicolumn{1}{p{0.5cm}}{\centering \xmark}
&\multicolumn{1}{p{0.5cm}}{\centering \cmark}
&\multicolumn{1}{p{0.5cm}}{\centering  \cmark}
&\multicolumn{1}{p{0.5cm}}{\centering \xmark}
&\multicolumn{1}{p{0.5cm}}{\centering \cmark}
&\multicolumn{1}{p{0.5cm}}{\centering \cmark}
&\multicolumn{1}{p{0.5cm}}{\centering \cmark}
&\multicolumn{1}{p{0.5cm}}{\centering \cmark}
&\multicolumn{1}{p{0.5cm}}{\centering  \xmark}
&\multicolumn{1}{p{0.5cm}}{\centering \xmark}
&\multicolumn{1}{p{0.5cm}}{\centering -}
&\multicolumn{1}{p{0.5cm}}{\centering \xmark}
&\multicolumn{1}{p{0.5cm}}{\centering  \cmark}
&\multicolumn{1}{p{0.5cm}}{\centering \xmark}
&\multicolumn{1}{p{0.5cm}}{\centering \xmark}
\\[6pt]\\

\multicolumn{1}{p{1cm}}{\centering Nasr and Fahmy\cite{nasr2017millimeter}}
&\multicolumn{1}{p{2cm}}{\centering Presented the comparisons of mesh and star topology for mmWave wireless backhauling for 5G}
&\multicolumn{1}{p{0.5cm}}{\centering \cmark}
&\multicolumn{1}{p{0.5cm}}{\centering \xmark}
&\multicolumn{1}{p{0.5cm}}{\centering  \cmark}
&\multicolumn{1}{p{0.5cm}}{\centering \cmark}
&\multicolumn{1}{p{0.5cm}}{\centering \cmark}
&\multicolumn{1}{p{0.5cm}}{\centering \xmark}
&\multicolumn{1}{p{0.5cm}}{\centering \xmark}
&\multicolumn{1}{p{0.5cm}}{\centering \xmark}
&\multicolumn{1}{p{0.5cm}}{\centering  \xmark}
&\multicolumn{1}{p{0.5cm}}{\centering \xmark}
&\multicolumn{1}{p{0.5cm}}{\centering \xmark}
&\multicolumn{1}{p{0.5cm}}{\centering \cmark}
&\multicolumn{1}{p{0.5cm}}{\centering \xmark}
&\multicolumn{1}{p{0.5cm}}{\centering \xmark}
&\multicolumn{1}{p{0.5cm}}{\centering  \xmark}
&\multicolumn{1}{p{0.5cm}}{\centering \xmark}
&\multicolumn{1}{p{0.5cm}}{\centering \xmark}
&\multicolumn{1}{p{0.5cm}}{\centering \cmark}
&\multicolumn{1}{p{0.5cm}}{\centering  \xmark}
&\multicolumn{1}{p{0.5cm}}{\centering \xmark}
&\multicolumn{1}{p{0.5cm}}{\centering \xmark}
\\[6pt]\\

\multicolumn{1}{p{1cm}}{\centering Destino et al.\cite{destino2017system}}
&\multicolumn{1}{p{2cm}}{\centering Considered the RF component requirements for mmWave mobile backhaul Transceivers}
&\multicolumn{1}{p{0.5cm}}{\centering \xmark}
&\multicolumn{1}{p{0.5cm}}{\centering \cmark}
&\multicolumn{1}{p{0.5cm}}{\centering  \xmark}
&\multicolumn{1}{p{0.5cm}}{\centering \cmark}
&\multicolumn{1}{p{0.5cm}}{\centering \xmark}
&\multicolumn{1}{p{0.5cm}}{\centering \xmark}
&\multicolumn{1}{p{0.5cm}}{\centering \cmark}
&\multicolumn{1}{p{0.5cm}}{\centering \xmark}
&\multicolumn{1}{p{0.5cm}}{\centering  \xmark}
&\multicolumn{1}{p{0.5cm}}{\centering \xmark}
&\multicolumn{1}{p{0.5cm}}{\centering \xmark}
&\multicolumn{1}{p{0.5cm}}{\centering \cmark}
&\multicolumn{1}{p{0.5cm}}{\centering \xmark}
&\multicolumn{1}{p{0.5cm}}{\centering \xmark}
&\multicolumn{1}{p{0.5cm}}{\centering  \xmark}
&\multicolumn{1}{p{0.5cm}}{\centering \xmark}
&\multicolumn{1}{p{0.5cm}}{\centering \cmark}
&\multicolumn{1}{p{0.5cm}}{\centering \xmark}
&\multicolumn{1}{p{0.5cm}}{\centering  \xmark}
&\multicolumn{1}{p{0.5cm}}{\centering \xmark}
&\multicolumn{1}{p{0.5cm}}{\centering \xmark}
\\[6pt]\\

\multicolumn{1}{p{1cm}}{\centering Hu and Blough\cite{hu2017relay}}
&\multicolumn{1}{p{2cm}}{\centering Proposed a new mmWave backhaul network architecture and algorithm for relay selection and scheduling}
&\multicolumn{1}{p{0.5cm}}{\centering \cmark}
&\multicolumn{1}{p{0.5cm}}{\centering \cmark}
&\multicolumn{1}{p{0.5cm}}{\centering  \xmark}
&\multicolumn{1}{p{0.5cm}}{\centering \cmark}
&\multicolumn{1}{p{0.5cm}}{\centering \cmark}
&\multicolumn{1}{p{0.5cm}}{\centering \xmark}
&\multicolumn{1}{p{0.5cm}}{\centering \xmark}
&\multicolumn{1}{p{0.5cm}}{\centering \cmark}
&\multicolumn{1}{p{0.5cm}}{\centering  \xmark}
&\multicolumn{1}{p{0.5cm}}{\centering \xmark}
&\multicolumn{1}{p{0.5cm}}{\centering \xmark}
&\multicolumn{1}{p{0.5cm}}{\centering \cmark}
&\multicolumn{1}{p{0.5cm}}{\centering \cmark}
&\multicolumn{1}{p{0.5cm}}{\centering \xmark}
&\multicolumn{1}{p{0.5cm}}{\centering  \xmark}
&\multicolumn{1}{p{0.5cm}}{\centering \cmark}
&\multicolumn{1}{p{0.5cm}}{\centering \xmark}
&\multicolumn{1}{p{0.5cm}}{\centering \cmark}
&\multicolumn{1}{p{0.5cm}}{\centering  \cmark}
&\multicolumn{1}{p{0.5cm}}{\centering \xmark}
&\multicolumn{1}{p{0.5cm}}{\centering \xmark}
\\[6pt]\\

\multicolumn{1}{p{1cm}}{\centering Magne et al.\cite{magne2018millimeter}}
&\multicolumn{1}{p{2cm}}{\centering Focused on the ownership cost and flexibility of the mmWave Point to Multipoint in backhaul}
&\multicolumn{1}{p{0.5cm}}{\centering \xmark}
&\multicolumn{1}{p{0.5cm}}{\centering \cmark}
&\multicolumn{1}{p{0.5cm}}{\centering  \cmark}
&\multicolumn{1}{p{0.5cm}}{\centering \cmark}
&\multicolumn{1}{p{0.5cm}}{\centering \cmark}
&\multicolumn{1}{p{0.5cm}}{\centering \xmark}
&\multicolumn{1}{p{0.5cm}}{\centering \cmark}
&\multicolumn{1}{p{0.5cm}}{\centering \cmark}
&\multicolumn{1}{p{0.5cm}}{\centering  \xmark}
&\multicolumn{1}{p{0.5cm}}{\centering \xmark}
&\multicolumn{1}{p{0.5cm}}{\centering \xmark}
&\multicolumn{1}{p{0.5cm}}{\centering \cmark}
&\multicolumn{1}{p{0.5cm}}{\centering \xmark}
&\multicolumn{1}{p{0.5cm}}{\centering \xmark}
&\multicolumn{1}{p{0.5cm}}{\centering  \xmark}
&\multicolumn{1}{p{0.5cm}}{\centering \xmark}
&\multicolumn{1}{p{0.5cm}}{\centering \xmark}
&\multicolumn{1}{p{0.5cm}}{\centering \xmark}
&\multicolumn{1}{p{0.5cm}}{\centering  \xmark}
&\multicolumn{1}{p{0.5cm}}{\centering \cmark}
&\multicolumn{1}{p{0.5cm}}{\centering \cmark}
\\[6pt]\\

\multicolumn{1}{p{1cm}}{\centering Han et al.\cite{han2018achievable}}
&\multicolumn{1}{p{2cm}}{\centering Emphasized on the adoption of  in-band full-duplex with mmWave-based backhaul network}
&\multicolumn{1}{p{0.5cm}}{\centering \xmark}
&\multicolumn{1}{p{0.5cm}}{\centering \cmark}
&\multicolumn{1}{p{0.5cm}}{\centering  \xmark}
&\multicolumn{1}{p{0.5cm}}{\centering \xmark}
&\multicolumn{1}{p{0.5cm}}{\centering \xmark}
&\multicolumn{1}{p{0.5cm}}{\centering \xmark}
&\multicolumn{1}{p{0.5cm}}{\centering \xmark}
&\multicolumn{1}{p{0.5cm}}{\centering \xmark}
&\multicolumn{1}{p{0.5cm}}{\centering  \xmark}
&\multicolumn{1}{p{0.5cm}}{\centering \xmark}
&\multicolumn{1}{p{0.5cm}}{\centering \xmark}
&\multicolumn{1}{p{0.5cm}}{\centering \cmark}
&\multicolumn{1}{p{0.5cm}}{\centering \xmark}
&\multicolumn{1}{p{0.5cm}}{\centering \xmark}
&\multicolumn{1}{p{0.5cm}}{\centering  \xmark}
&\multicolumn{1}{p{0.5cm}}{\centering \xmark}
&\multicolumn{1}{p{0.5cm}}{\centering \xmark}
&\multicolumn{1}{p{0.5cm}}{\centering \xmark}
&\multicolumn{1}{p{0.5cm}}{\centering  \xmark}
&\multicolumn{1}{p{0.5cm}}{\centering \xmark}
&\multicolumn{1}{p{0.5cm}}{\centering \xmark}
\\[6pt]\\

\multicolumn{1}{p{1cm}}{\centering Koslowski et al.\cite{koslowski2018meshed}}
&\multicolumn{1}{p{2cm}}{\centering Focused on the link overloading or link failure}
&\multicolumn{1}{p{0.5cm}}{\centering \cmark}
&\multicolumn{1}{p{0.5cm}}{\centering \cmark}
&\multicolumn{1}{p{0.5cm}}{\centering  \cmark}
&\multicolumn{1}{p{0.5cm}}{\centering \cmark}
&\multicolumn{1}{p{0.5cm}}{\centering \cmark}
&\multicolumn{1}{p{0.5cm}}{\centering \xmark}
&\multicolumn{1}{p{0.5cm}}{\centering \cmark}
&\multicolumn{1}{p{0.5cm}}{\centering \xmark}
&\multicolumn{1}{p{0.5cm}}{\centering  \xmark}
&\multicolumn{1}{p{0.5cm}}{\centering \xmark}
&\multicolumn{1}{p{0.5cm}}{\centering \xmark}
&\multicolumn{1}{p{0.5cm}}{\centering \cmark}
&\multicolumn{1}{p{0.5cm}}{\centering \cmark}
&\multicolumn{1}{p{0.5cm}}{\centering \xmark}
&\multicolumn{1}{p{0.5cm}}{\centering  \xmark}
&\multicolumn{1}{p{0.5cm}}{\centering \xmark}
&\multicolumn{1}{p{0.5cm}}{\centering \xmark}
&\multicolumn{1}{p{0.5cm}}{\centering \xmark}
&\multicolumn{1}{p{0.5cm}}{\centering  \xmark}
&\multicolumn{1}{p{0.5cm}}{\centering \cmark}
&\multicolumn{1}{p{0.5cm}}{\centering \xmark}
\\[6pt]\\

\multicolumn{1}{p{1cm}}{Sahoo et al.\cite{sahoo2017millimeter}}
&\multicolumn{1}{p{2cm}}{\centering Proposed a frame reconfiguration scheme for traffic management and scheduling algorithm }
&\multicolumn{1}{p{0.5cm}}{\centering \cmark}
&\multicolumn{1}{p{0.5cm}}{\centering \cmark}
&\multicolumn{1}{p{0.5cm}}{\centering  \xmark}
&\multicolumn{1}{p{0.5cm}}{\centering \xmark}
&\multicolumn{1}{p{0.5cm}}{\centering \cmark}
&\multicolumn{1}{p{0.5cm}}{\centering \xmark}
&\multicolumn{1}{p{0.5cm}}{\centering \xmark}
&\multicolumn{1}{p{0.5cm}}{\centering \xmark}
&\multicolumn{1}{p{0.5cm}}{\centering  \xmark}
&\multicolumn{1}{p{0.5cm}}{\centering \xmark}
&\multicolumn{1}{p{0.5cm}}{\centering \xmark}
&\multicolumn{1}{p{0.5cm}}{\centering \cmark}
&\multicolumn{1}{p{0.5cm}}{\centering \cmark}
&\multicolumn{1}{p{0.5cm}}{\centering \xmark}
&\multicolumn{1}{p{0.5cm}}{\centering  \xmark}
&\multicolumn{1}{p{0.5cm}}{\centering -}
&\multicolumn{1}{p{0.5cm}}{\centering \cmark}
&\multicolumn{1}{p{0.5cm}}{\centering \cmark}
&\multicolumn{1}{p{0.5cm}}{\centering  \cmark}
&\multicolumn{1}{p{0.5cm}}{\centering \cmark}
&\multicolumn{1}{p{0.5cm}}{\centering \xmark}
\\[6pt]\\

\multicolumn{1}{p{1cm}}{Feng et al.\cite{feng2016millimetre}}
&\multicolumn{1}{p{2cm}}{\centering Proposed a 5G mmWave backhaul framework and discussed the challenges and design issues }
&\multicolumn{1}{p{0.5cm}}{\centering \cmark}
&\multicolumn{1}{p{0.5cm}}{\centering \cmark}
&\multicolumn{1}{p{0.5cm}}{\centering  -}
&\multicolumn{1}{p{0.5cm}}{\centering \cmark}
&\multicolumn{1}{p{0.5cm}}{\centering \cmark}
&\multicolumn{1}{p{0.5cm}}{\centering \cmark}
&\multicolumn{1}{p{0.5cm}}{\centering \xmark}
&\multicolumn{1}{p{0.5cm}}{\centering \cmark}
&\multicolumn{1}{p{0.5cm}}{\centering  \xmark}
&\multicolumn{1}{p{0.5cm}}{\centering \xmark}
&\multicolumn{1}{p{0.5cm}}{\centering \xmark}
&\multicolumn{1}{p{0.5cm}}{\centering \cmark}
&\multicolumn{1}{p{0.5cm}}{\centering \xmark}
&\multicolumn{1}{p{0.5cm}}{\centering \xmark}
&\multicolumn{1}{p{0.5cm}}{\centering  \xmark}
&\multicolumn{1}{p{0.5cm}}{\centering \cmark}
&\multicolumn{1}{p{0.5cm}}{\centering \cmark}
&\multicolumn{1}{p{0.5cm}}{\centering \cmark}
&\multicolumn{1}{p{0.5cm}}{\centering  \cmark}
&\multicolumn{1}{p{0.5cm}}{\centering \cmark}
&\multicolumn{1}{p{0.5cm}}{\centering \xmark}
\\[6pt]\\

\multicolumn{1}{p{1cm}}{Pateromichelakis et al.\cite{pateromichelakis2017slice}}
&\multicolumn{1}{p{2cm}}{\centering Emphasized on the path selection and scheduling, routing problems in backhaul and give a multi-tenant framework}
&\multicolumn{1}{p{0.5cm}}{\centering \cmark}
&\multicolumn{1}{p{0.5cm}}{\centering \cmark}
&\multicolumn{1}{p{0.5cm}}{\centering  \xmark}
&\multicolumn{1}{p{0.5cm}}{\centering \cmark}
&\multicolumn{1}{p{0.5cm}}{\centering \cmark}
&\multicolumn{1}{p{0.5cm}}{\centering \xmark}
&\multicolumn{1}{p{0.5cm}}{\centering \xmark}
&\multicolumn{1}{p{0.5cm}}{\centering \xmark}
&\multicolumn{1}{p{0.5cm}}{\centering  \xmark}
&\multicolumn{1}{p{0.5cm}}{\centering \cmark}
&\multicolumn{1}{p{0.5cm}}{\centering \xmark}
&\multicolumn{1}{p{0.5cm}}{\centering \cmark}
&\multicolumn{1}{p{0.5cm}}{\centering \xmark}
&\multicolumn{1}{p{0.5cm}}{\centering \xmark}
&\multicolumn{1}{p{0.5cm}}{\centering  \xmark}
&\multicolumn{1}{p{0.5cm}}{\centering \xmark}
&\multicolumn{1}{p{0.5cm}}{\centering \xmark}
&\multicolumn{1}{p{0.5cm}}{\centering \cmark}
&\multicolumn{1}{p{0.5cm}}{\centering  \cmark}
&\multicolumn{1}{p{0.5cm}}{\centering \xmark}
&\multicolumn{1}{p{0.5cm}}{\centering \xmark}
\\[6pt]\\

\hline

\end{longtable}%}
\end{center}
\end{landscape}

Han et al.\cite{han2018achievable} emphasized the adoption of in-band full-duplex with a mmWave-based backhaul network. Link overloading or link failure issues are highlighted by Koslowski et al.\cite{koslowski2018meshed}. Sahoo et al.\cite{sahoo2017millimeter} provided a frame reconfiguration scheme for traffic management, and proposed a scheduling algorithm for power allocation. Feng et al.\cite{feng2016millimetre} gave a 5G mmWave backhaul framework, and discussed the challenges and design issues in mmWave backhaul. Pateromichelakis et al.\cite{pateromichelakis2017slice} emphasized the path selection, scheduling and routing problem in backhaul and presented a multi-tenant framework. The authors described dynamic on-demand slice requirements, and to handle such requirements, an SDN-based local coordinator is suggested. The prominent solutions based on mmWave are presented in Table~\ref{Tablemmwave}.

\section{Backhaul Security}
The excessive traffic leads to various security threats in the mobile backhaul networks. The emerging data and users over communication networks in 5G require security solutions which leverage the facility of traffic movements and sustainable networks without affecting the network performance. Various potential security solutions for the mobile backhaul are discussed in Table~\ref{Tablesecurity}.  Sharma et al.\cite{sharma2018secure} proposed a key exchange and authentication protocol for the handover in 5G mobile Xhaul networks against the eavesdropping and DoS attacks. The authors considered some significant security requirements such as mutual authentication, key exchange, perfect forward secrecy, and privacy for designing handover protocol. Liyanage et al.\cite{liyanage2014case} discussed the security issues on LTE-backhaul like DoS, unwanted communications via VoIP, and distributions of viruses. The authors discussed the TLS/SSL-based, IPsec tunnel mode, and IPsec BEET mode-based security solutions for the LTE-backhaul networks. The authors emphasized security issues in the traditional networks and suggested the futuristic solutions through secure GWs, firewalls, and IPS.

Rohlik and Vanek\cite{rohlik2013new} focused on the femtocell backhaul security and suggested IPsec tunnel security solutions. The authors discussed the applicability of transport layer security (TLS) and Datagram TLS (DTLS) protocols. Furthermore, Namal et al. \cite{namal2011securing} emphasized the mobility and security issues in femtocell and backhaul. The authors modified the 3GPP femtocell architecture protocol stack. Borgaonkar et al. \cite{borgaonkar2011security} focused on the security issues in the femtocell. Liyanage and Gurtov \cite{liyanage2012secured} emphasized on the Virtual Private Network (VPN) architectures for LTE backhaul. The authors integrated the Host Identity Protocol (HIP) and Internet Key exchange version 2 (IKEv2) in the IPsec. The authors presented the strength of the architecture against flooding and DoS attacks. Liyanage et al. \cite{liyanage2015security} focused on the SDN security and proposed security architecture based on the multi-tier approach. The proposed architecture provides HIP-based secure communication channels, policy-based communications, security management and monitoring, and synchronized network.

Chen et al. \cite{chen2016software} surveyed the security issues in software-defined mobile networks and classified the data, application, and control plane attack. The authors considered the STRIDE-based threat classification and their countermeasures. Clavister Solution Series\cite{clavister} focused on network attacks and the standards-based solution to secure LTE backhaul. Juniper Networks\cite{juniper} focused on the LTE backhaul requirements. Yaniv Balmas\cite{balmas2013mobile} discussed mobile network security and suggested the Deep Packet Inspection, VOIP and DNS Protections as potential solutions. Oracle Solutions\cite{oracle} considered the heterogeneous network and backhaul security for heterogeneous networks. SDN and Network Functions Virtualization (NFV) can enhance the performance, flexibility, and scalability of telecommunication networks. Despite these attempts, adequate solutions are still required to meet the security requirements of 5G mobile backhaul networks that should not pose any other complexities.

\begin{landscape}
\begin{center}
\fontsize{7}{8}\selectfont
\setlength\LTleft{10pt}            % default: \parindent
\setlength\LTright{0pt}
%\begin{longtable}{llllllllllll}
\begin{longtable}{@{\extracolsep{\fill}}*{12}{c}}
\caption{A comparison of existing security solutions for 5G mobile backhaul Networks.(R1:Standard Security Adoption, R2:Security Implementation cost consideration, R3:VPN-based traffic control, R4:Secure mobility management/Handover Security, R5:Tunnel Based Security, R6:De-centralized Security, R7:Software-based Security Solution)}\label{Tablesecurity} \\
\hline
\\[6pt]\\
\multicolumn{1}{p{1cm}}{\centering \textbf{Scheme}}
&\multicolumn{1}{p{2cm}}{\centering \textbf{Main Contributions}}
&\multicolumn{1}{p{2cm}}{\centering \textbf{Types of Attacks consider} }
&\multicolumn{1}{p{2cm}}{\centering \textbf{Types of Security Requirements consider}}
&\multicolumn{1}{p{2cm}}{\centering \textbf{Security Solutions}}
&\multicolumn{1}{p{0.5cm}}{\centering \textbf{R1}}
&\multicolumn{1}{p{0.5cm}}{\centering \textbf{R2}}
&\multicolumn{1}{p{0.5cm}}{\centering \textbf{R3}}
&\multicolumn{1}{p{0.5cm}}{\centering \textbf{R4}}
&\multicolumn{1}{p{0.5cm}}{\centering \textbf{R5}}
&\multicolumn{1}{p{0.5cm}}{\centering \textbf{R6}}
&\multicolumn{1}{p{0.5cm}}{\centering \textbf{R7}}

\\[6pt]\\
\hline \\%data entry for gu

\endfirsthead

\multicolumn{12}{c}%
{{\bfseries \tablename\ \thetable{} -- continued from previous page}} \\
\hline
\\[6pt]\\
\multicolumn{1}{p{1cm}}{\centering \textbf{Scheme}}
&\multicolumn{1}{p{2cm}}{\centering \textbf{Main Contributions}}
&\multicolumn{1}{p{2cm}}{\centering \textbf{Types of Attacks consider} }
&\multicolumn{1}{p{2cm}}{\centering \textbf{Types of Security Requirements consider}}
&\multicolumn{1}{p{2cm}}{\centering \textbf{Security Solutions}}
&\multicolumn{1}{p{0.5cm}}{\centering \textbf{R1}}
&\multicolumn{1}{p{0.5cm}}{\centering \textbf{R2}}
&\multicolumn{1}{p{0.5cm}}{\centering \textbf{R3}}
&\multicolumn{1}{p{0.5cm}}{\centering \textbf{R4}}
&\multicolumn{1}{p{0.5cm}}{\centering \textbf{R5}}
&\multicolumn{1}{p{0.5cm}}{\centering \textbf{R6}}
&\multicolumn{1}{p{0.5cm}}{\centering \textbf{R7}}
\\[6pt]\\
\hline \\
\endhead

\hline \multicolumn{12}{l}{{Continued on next page}} \\

\endfoot

\endlastfoot
\multicolumn{1}{p{1cm}}{\centering Sharma et al.\cite{sharma2018secure}}
&\multicolumn{1}{p{2cm}}{Proposed a key exchange and authentication
protocol for the handover in 5G mobile Xhaul networks}
&\multicolumn{1}{p{2cm}}{Eavesdropping, DoS}
&\multicolumn{1}{p{2cm}}{Mutual authentication, key exchange, perfect forward secrecy, privacy}
&\multicolumn{1}{p{2cm}}{Secure Handover Protocol}
&\multicolumn{1}{p{0.5cm}}{\centering \xmark}
&\multicolumn{1}{p{0.5cm}}{\centering \cmark}
&\multicolumn{1}{p{0.5cm}}{\centering \xmark}
&\multicolumn{1}{p{0.5cm}}{\centering \cmark}
&\multicolumn{1}{p{0.5cm}}{\centering \xmark}
&\multicolumn{1}{p{0.5cm}}{\centering  \xmark}
&\multicolumn{1}{p{0.5cm}}{\centering \xmark}
\\[6pt]\\

\multicolumn{1}{p{1cm}}{\centering Liyanage et al.\cite{liyanage2014case}}
&\multicolumn{1}{p{2cm}}{Discussed security issues on LTE-backhaul}
&\multicolumn{1}{p{2cm}}{DoS, unwanted communications via VoIP, Distributions of viruses}
&\multicolumn{1}{p{2cm}}{Payload encryption, authentication}
&\multicolumn{1}{p{2cm}}{Deep Packet Inspection, IPsec traffic transportation, firewall and security architectures-TLS/SSL-Based, IPsec Tunnel Mode, IPsec BEET Mode}
&\multicolumn{1}{p{0.5cm}}{\centering \cmark}
&\multicolumn{1}{p{0.5cm}}{\centering \cmark}
&\multicolumn{1}{p{0.5cm}}{\centering \cmark}
&\multicolumn{1}{p{0.5cm}}{\centering \xmark}
&\multicolumn{1}{p{0.5cm}}{\centering \cmark}
&\multicolumn{1}{p{0.5cm}}{\centering  \cmark}
&\multicolumn{1}{p{0.5cm}}{\centering \cmark}
\\[6pt]\\

\multicolumn{1}{p{1cm}}{\centering Rohlik and Vanek\cite{rohlik2013new}}
&\multicolumn{1}{p{2cm}}{Focused on the Femtocell backhaul security and  suggest IPsec tunnel security}
&\multicolumn{1}{p{2cm}}{\centering-}
&\multicolumn{1}{p{2cm}}{Mutual authentication, privacy}
&\multicolumn{1}{p{2cm}}{Transport Layer Security, IKEv2 protocol}
&\multicolumn{1}{p{0.5cm}}{\centering \cmark}
&\multicolumn{1}{p{0.5cm}}{\centering \xmark}
&\multicolumn{1}{p{0.5cm}}{\centering \xmark}
&\multicolumn{1}{p{0.5cm}}{\centering \xmark}
&\multicolumn{1}{p{0.5cm}}{\centering \cmark}
&\multicolumn{1}{p{0.5cm}}{\centering  \xmark}
&\multicolumn{1}{p{0.5cm}}{\centering \xmark}
\\[6pt]\\

\multicolumn{1}{p{1cm}}{\centering Namal et al. \cite{namal2011securing}}
&\multicolumn{1}{p{2cm}}{Emphasized on the mobility and security issues in femtocell and backhaul}
&\multicolumn{1}{p{2cm}}{DoS, Distributed DoS}
&\multicolumn{1}{p{2cm}}{Authentication, service registration, identity
Verification, confidentiality, Integrity}
&\multicolumn{1}{p{2cm}}{Host Identity Protocol, Encapsulating Security Payload, EAP-AKA}
&\multicolumn{1}{p{0.5cm}}{\centering \cmark}
&\multicolumn{1}{p{0.5cm}}{\centering \xmark}
&\multicolumn{1}{p{0.5cm}}{\centering \xmark}
&\multicolumn{1}{p{0.5cm}}{\centering \cmark}
&\multicolumn{1}{p{0.5cm}}{\centering \cmark}
&\multicolumn{1}{p{0.5cm}}{\centering  \cmark}
&\multicolumn{1}{p{0.5cm}}{\centering \xmark}
\\[6pt]\\

\multicolumn{1}{p{1cm}}{\centering Liyanage and Gurtov
\cite{liyanage2012secured}}
&\multicolumn{1}{p{2cm}}{Emphasized on the VPN architectures for LTE-backhaul}
&\multicolumn{1}{p{2cm}}{DoS, DDoS and TCP reset attacks}
&\multicolumn{1}{p{2cm}}{User Authentication and authorization, Payload Encryption, privacy, Attack Prevention}
&\multicolumn{1}{p{2cm}}{IP security, host identity protocol}
&\multicolumn{1}{p{0.5cm}}{\centering \cmark}
&\multicolumn{1}{p{0.5cm}}{\centering \cmark}
&\multicolumn{1}{p{0.5cm}}{\centering \cmark}
&\multicolumn{1}{p{0.5cm}}{\centering \cmark}
&\multicolumn{1}{p{0.5cm}}{\centering \cmark}
&\multicolumn{1}{p{0.5cm}}{\centering  \xmark}
&\multicolumn{1}{p{0.5cm}}{\centering \xmark}
\\[6pt]\\

\multicolumn{1}{p{1cm}}{\centering Liyanage et al. \cite{liyanage2015security}}
&\multicolumn{1}{p{2cm}}{Focused on the SDN security and proposed a security architecture}
&\multicolumn{1}{p{2cm}}{Spoofing and DoS attacks}
&\multicolumn{1}{p{2cm}}{Authentication}
&\multicolumn{1}{p{2cm}}{Host Identity Protocol based secure tunneling}
&\multicolumn{1}{p{0.5cm}}{\centering \cmark}
&\multicolumn{1}{p{0.5cm}}{\centering \xmark}
&\multicolumn{1}{p{0.5cm}}{\centering \xmark}
&\multicolumn{1}{p{0.5cm}}{\centering \cmark}
&\multicolumn{1}{p{0.5cm}}{\centering \cmark}
&\multicolumn{1}{p{0.5cm}}{\centering  \xmark}
&\multicolumn{1}{p{0.5cm}}{\centering \cmark}
\\[6pt]\\

\multicolumn{1}{p{1cm}}{\centering Chen et al. \cite{chen2016software}}
&\multicolumn{1}{p{2cm}}{Surveyed the security issues in software defined mobile networks}
&\multicolumn{1}{p{2cm}}{Viruses, replay attack, topological poison, Decoy attacks, tampering-related attack,  DoS, botnet, data, application and control plane attacks}
&\multicolumn{1}{p{2cm}}{Authentication, encryption, integrity, availability}
&\multicolumn{1}{p{2cm}}{STRIDE-based threat classification and their countermeasures}
&\multicolumn{1}{p{0.5cm}}{\centering \cmark}
&\multicolumn{1}{p{0.5cm}}{\centering \cmark}
&\multicolumn{1}{p{0.5cm}}{\centering \xmark}
&\multicolumn{1}{p{0.5cm}}{\centering \cmark}
&\multicolumn{1}{p{0.5cm}}{\centering \cmark}
&\multicolumn{1}{p{0.5cm}}{\centering  \xmark}
&\multicolumn{1}{p{0.5cm}}{\centering \cmark}
\\[6pt]\\

\multicolumn{1}{p{1cm}}{\centering Clavister Solution Series\cite{clavister}}
&\multicolumn{1}{p{2cm}}{Focused on the standards-based solution to secure LTE backhaul}
&\multicolumn{1}{p{2cm}}{Network attacks}
&\multicolumn{1}{p{2cm}}{Availability, Subscriber data Security}
&\multicolumn{1}{p{2cm}}{IPsec, Clavister’s mobile backhaul solution}
&\multicolumn{1}{p{0.5cm}}{\centering \cmark}
&\multicolumn{1}{p{0.5cm}}{\centering \cmark}
&\multicolumn{1}{p{0.5cm}}{\centering \xmark}
&\multicolumn{1}{p{0.5cm}}{\centering \xmark}
&\multicolumn{1}{p{0.5cm}}{\centering \cmark}
&\multicolumn{1}{p{0.5cm}}{\centering  \xmark}
&\multicolumn{1}{p{0.5cm}}{\centering \cmark}
\\[6pt]\\

\multicolumn{1}{p{1cm}}{Juniper Networks\cite{juniper}}
&\multicolumn{1}{p{2cm}}{Focused on the LTE backhaul requirements}
&\multicolumn{1}{p{2cm}}{DoS, Man-in-Middle Attack}
&\multicolumn{1}{p{2cm}}{Availability, Encryption}
&\multicolumn{1}{p{2cm}}{IPSec, Firewall}
&\multicolumn{1}{p{0.5cm}}{\centering \cmark}
&\multicolumn{1}{p{0.5cm}}{\centering \cmark}
&\multicolumn{1}{p{0.5cm}}{\centering \xmark}
&\multicolumn{1}{p{0.5cm}}{\centering -}
&\multicolumn{1}{p{0.5cm}}{\centering \cmark}
&\multicolumn{1}{p{0.5cm}}{\centering  \xmark}
&\multicolumn{1}{p{0.5cm}}{\centering \cmark}
\\[6pt]\\

\multicolumn{1}{p{1cm}}{Liyanage et al. \cite{liyanage2016novel}}
&\multicolumn{1}{p{2cm}}{Proposed a secure VPN-based traffic architectures for LTE-backhaul}
&\multicolumn{1}{p{2cm}}{IP-based attacks, DoS, Replay attacks, eavesdropping, Spoofing}
&\multicolumn{1}{p{2cm}}{Authentication, authorization, payload encryption, privacy protection, and IP-based attack
prevention}
&\multicolumn{1}{p{2cm}}{HIP, Internet key exchange version 2, IPsec}
&\multicolumn{1}{p{0.5cm}}{\centering \cmark}
&\multicolumn{1}{p{0.5cm}}{\centering \cmark}
&\multicolumn{1}{p{0.5cm}}{\centering \cmark}
&\multicolumn{1}{p{0.5cm}}{\centering \cmark}
&\multicolumn{1}{p{0.5cm}}{\centering \cmark}
&\multicolumn{1}{p{0.5cm}}{\centering  \cmark}
&\multicolumn{1}{p{0.5cm}}{\centering \xmark}
\\[6pt]\\

\multicolumn{1}{p{1cm}}{Yaniv Balmas\cite{balmas2013mobile}}
&\multicolumn{1}{p{2cm}}{Focused on the mobile network security}
&\multicolumn{1}{p{2cm}}{Zero-day attacks, blown attack, DoS, flooding attacks}
&\multicolumn{1}{p{2cm}}{Security risk assessment, authentication, privacy}
&\multicolumn{1}{p{2cm}}{Deep Packet Inspection, VOIP and DNS Protections}
&\multicolumn{1}{p{0.5cm}}{\centering \cmark}
&\multicolumn{1}{p{0.5cm}}{\centering \cmark}
&\multicolumn{1}{p{0.5cm}}{\centering \xmark}
&\multicolumn{1}{p{0.5cm}}{\centering \xmark}
&\multicolumn{1}{p{0.5cm}}{\centering \xmark}
&\multicolumn{1}{p{0.5cm}}{\centering  -}
&\multicolumn{1}{p{0.5cm}}{\centering \cmark}
\\[6pt]\\

\multicolumn{1}{p{1cm}}{Borgaonkar et al. \cite{borgaonkar2011security}}
&\multicolumn{1}{p{2cm}}{Emphasized on the security issues in femtocell}
&\multicolumn{1}{p{2cm}}{protocol attacks, privacy attacks}
&\multicolumn{1}{p{2cm}}{Mutual Authentication, encryption}
&\multicolumn{1}{p{2cm}}{IPsec tunnel, EAP-SIM, EAP-AKA}
&\multicolumn{1}{p{0.5cm}}{\centering \cmark}
&\multicolumn{1}{p{0.5cm}}{\centering \cmark}
&\multicolumn{1}{p{0.5cm}}{\centering \xmark}
&\multicolumn{1}{p{0.5cm}}{\centering \xmark}
&\multicolumn{1}{p{0.5cm}}{\centering \cmark}
&\multicolumn{1}{p{0.5cm}}{\centering  \xmark}
&\multicolumn{1}{p{0.5cm}}{\centering \cmark}
\\[6pt]\\

\multicolumn{1}{p{1cm}}{Oracle\cite{oracle}}
&\multicolumn{1}{p{2cm}}{Focused on the heterogeneous network and backhaul security}
&\multicolumn{1}{p{2cm}}{Malicious attacks, DoS and DDoS}
&\multicolumn{1}{p{2cm}}{Authentication, encryption, and control functions}
&\multicolumn{1}{p{2cm}}{Hardware-based  tunnel and encryption, IPSec}
&\multicolumn{1}{p{0.5cm}}{\centering \cmark}
&\multicolumn{1}{p{0.5cm}}{\centering \cmark}
&\multicolumn{1}{p{0.5cm}}{\centering \xmark}
&\multicolumn{1}{p{0.5cm}}{\centering \cmark}
&\multicolumn{1}{p{0.5cm}}{\centering \cmark}
&\multicolumn{1}{p{0.5cm}}{\centering  \xmark}
&\multicolumn{1}{p{0.5cm}}{\centering \xmark}
\\[6pt]\\

\hline

\end{longtable}%}
\end{center}
\end{landscape}

\subsection{Security Requirements for 5G Mobile Backhaul}
\begin{figure}[ht!]
\centering
\includegraphics[width=380px]{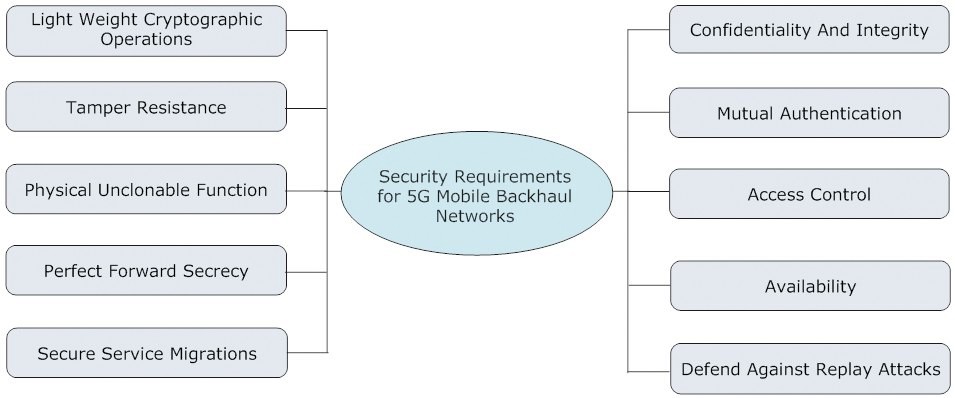}
\caption{The security requirements of 5G mobile backhaul.}
\label{requirements}
\end{figure}

The intense need for security mechanisms for trusted communication networks has facilitated various standards to incorporate feasible security solutions to achieve the network security requirements~\cite{fang2018security}. The security threats affect the privacy, integrity, and availability of the networks. The 5G backhaul security also needs an adequate security solution which can cover all the basic requirements. Some of the major security requirements for 5G mobile backhaul are shown in Fig.~\ref{requirements}.

\begin{itemize}
\item \textbf{Confidentiality and Integrity}: The confidentiality and integrity are the basic security requirements for the backhaul network. In the communication system, all of the control signals and transmitted user data can be attacked by malicious node through eavesdropping, modification, and man-in-middle attacks. The protection of such crucial data is necessary from these threats~\cite{prasad20183gpp}.

\item \textbf{Mutual Authentication}:  It is an important key element for defense against camouflage attacks. The 5G network must be verified and the direct communication between each device and the core network should be authenticated. In the 5G mobile backhaul, the terminal and hub should be mutually authenticated for utilization of links~\cite{basin2018formal}. The secure channel between device and network can be established through the secure key. Node authentication prevents unauthorized access to the backhaul network and ensures that the traffic is flowing between legitimate users~\cite{moreira2018cross}.

\item \textbf{Access Control}: In a 5G network, user’s personal profiles and other factors need a granular access control policy.  The granular access control policies help to prevent the acquisition of the information from the unauthorized users. The valid access controls help to defend against privacy and data transfer threats~\cite{ahmad2018overview}.

\item\textbf{Availability}:  It is an important factor in communication between group-based devices that ensures inter-device communication services for authorized users. The malicious and non-legitimate users affect the channel and service accessibility for the legitimate authorities. The service and channel availability can be secured by abolishing the privileges of an ineligible user~\cite{liyanage2018comprehensive}. The DoS and distributed DoS are major attacks which affect the availability of backhaul networks and should be mitigated.

\item \textbf{Perfect Forward Secrecy}: The
data in the backhaul networks must be protected by keys. The data between the core and the hub should be protected through session keys; therefore, it has to be assured that session keys will not be compromised even if the private keys of the server are compromised.

\item \textbf{Light-Weight Cryptographic Operations}:  In order to support confidentiality and integrity in the network, cryptographic operations are essential. The speed of cryptographic operations is important because it satisfies high-speed transmission rate and low latency. The light-weight cryptographic operations enhance the life of backhaul devices. Therefore, the security solutions should emphasize the light-weight-cryptographic operations in the 5G mobile backhaul to avoid degradation of the network performance in terms of delays and latency~\cite{kornaros2018hardware}.

\item \textbf{Tamper Resistance}: The master key used in the network can be leaked through physical tampering, which means that an attacker may be able to send sensitive information bypassing the access controls. To ensure the security of the codes or packages installed in the devices, tamper-resistant security modules can be adopted. These modules provide the physical protection to the cryptographic functions and prevent the compromise of secret keys~\cite{fisher2018enclosure}.

\item     \textbf{Defend Against Replay Attacks}: The freshness of keys and seeds are essential requirements for the secure communications. The replay attacks can use bogus requests for gaining access in the backhaul networks~\cite{sharma2018secure}. The freshness ensures the legitimacy of requests and prevents the delicacy of the authentication requests. The freshness can help to mitigate spoofing attacks as well as DoS attacks.

\item \textbf{Secure Service Migrations}: The services over the mobile networks can be migrated according to the requirements of various users~\cite{ma2018efficient}. The security of such services should maintain its security environment in case of migrations to another environment. The mobility management and handover management should be adopted to secure the service migrations in the mobile backhaul network.

\item \textbf{Physical Unclonable Function (PUF)}:  The PUF used as a unique identifier on the basis of physical variation in the semiconductor~\cite{jisis18-8-3-03}. PUF is a digital fingerprint that is unique to each IC with an output value that cannot be reverse-engineered. The PUF can be implanted at a very small hardware cost. The randomness of the PUF is based on environmental variations like temperature, supply voltage, and electromagnetic interference~\cite{tajik2019physical}. The PUF output is used as a unique key/secret to support cryptographic algorithms and services that include encryption/decryption, authentication, and digital signatures.
\end{itemize}

The protection of mobile backhaul against various threats incorporates multiple security add-ons. The protection against threats without affecting QoS, performance, and reliability of the network is a major concern~\cite{roman2018mobile}. These security requirements are indeed major considerations for trusted backhaul network solutions at every level of operations.

\subsection{Solutions and Technologies for Backhaul Security}
It is evident from the existing literature that the emerging technologies empower the communication networks and also open opportunities for diverse security attacks due to existing vulnerabilities and lack of security solutions~\cite{gupta2018security}. The traditional networks, 2G, 3G, and LTE have faced a lot of security issues and some of these security issues also exist in the 5G network. The attacks on the network by insider threats, external attackers or malicious nodes/users defame the performance of the backhaul and affect its availability. To provide a trusted network, various security solutions have been adopted in order to maintain the privacy and confidentiality of backhaul networks~\cite{yang2018blockonet}.

\begin{figure}[ht!]
\centering
\includegraphics[width=380px]{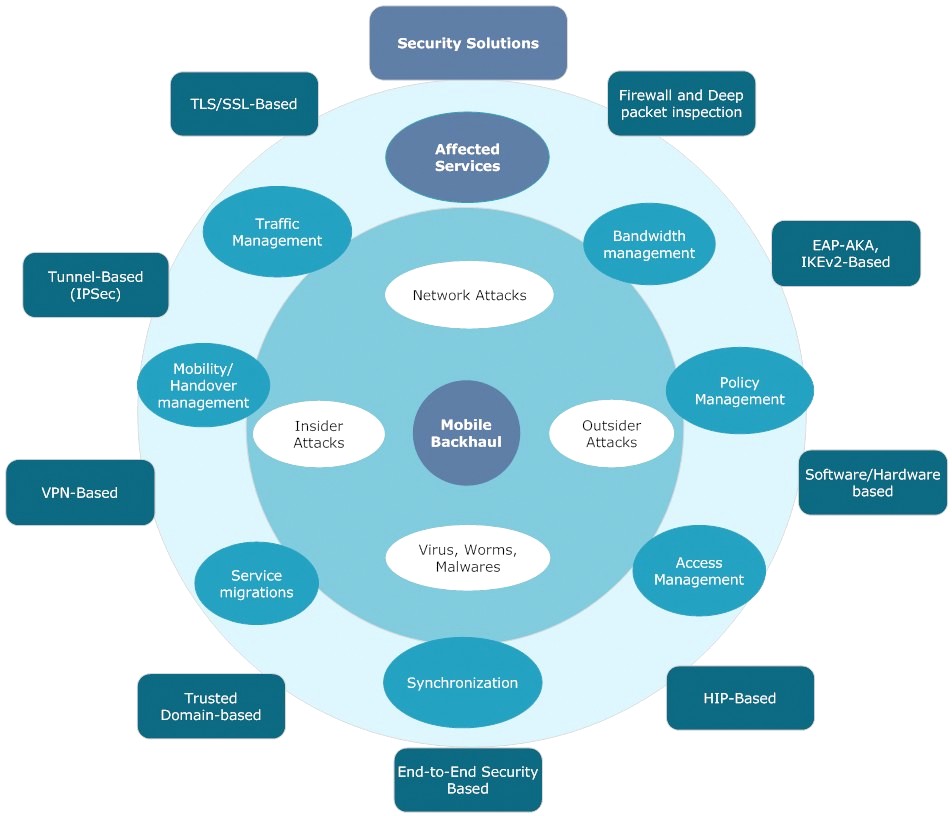}
\caption{The security solutions for mobile backhaul, attacks and affected services.}
\label{securitysolutions}
\end{figure}

The existing security solutions, as shown in Fig~\ref{securitysolutions}, emphasized on the trusted networks via control of physical site locations, ownership, and secure administrative controls. IPSec tunnels are used to secure IP traffic and provide backhaul protection against attackers~\cite{navarro2018integration}. Encapsulating Security Payload (ESP) are used for integrity, confidentiality and it also provides attack persistence~\cite{getschmann2018multi}. A malicious user can manipulate the public key and gain access to the secure gateways. To provide protection against such cases, the EAP-AKA can be a potential solution. The deep packet inspection, firewalls, and software/hardware-based security are also adaptable in the backhaul networks. TLS/SSL~\cite{eiriksson2018method} is integrated with the VPN-based traffic transportations. It is conclusive that the 5G backhaul security needs a dedicated security architecture which provides security at each level of communication and maintains the network performance~\cite{sharma2018resource}.

In 5G, the communication links between a base station and the associated mobile switching nodes are used to transmit the signals and data. The base stations are used for radio coverage and communications with mobile users. The mobile backhaul links can use microwave radio for better capacity and coverage~\cite{chang20185g}. The emerging technologies such as Ethernet-based solutions, mmWave radios, and satellite-backhaul solutions are responsible for the rapid expansion of radio coverage. In the earlier era of communication networks, mobile backhaul established a strong pillar with the adequate infrastructures. But emerging demands of networks such as high speed, low latency, and maximum throughput places new requirements of mobile backhaul networks.

\begin{figure}[ht!]
\centering
\includegraphics[width=420px]{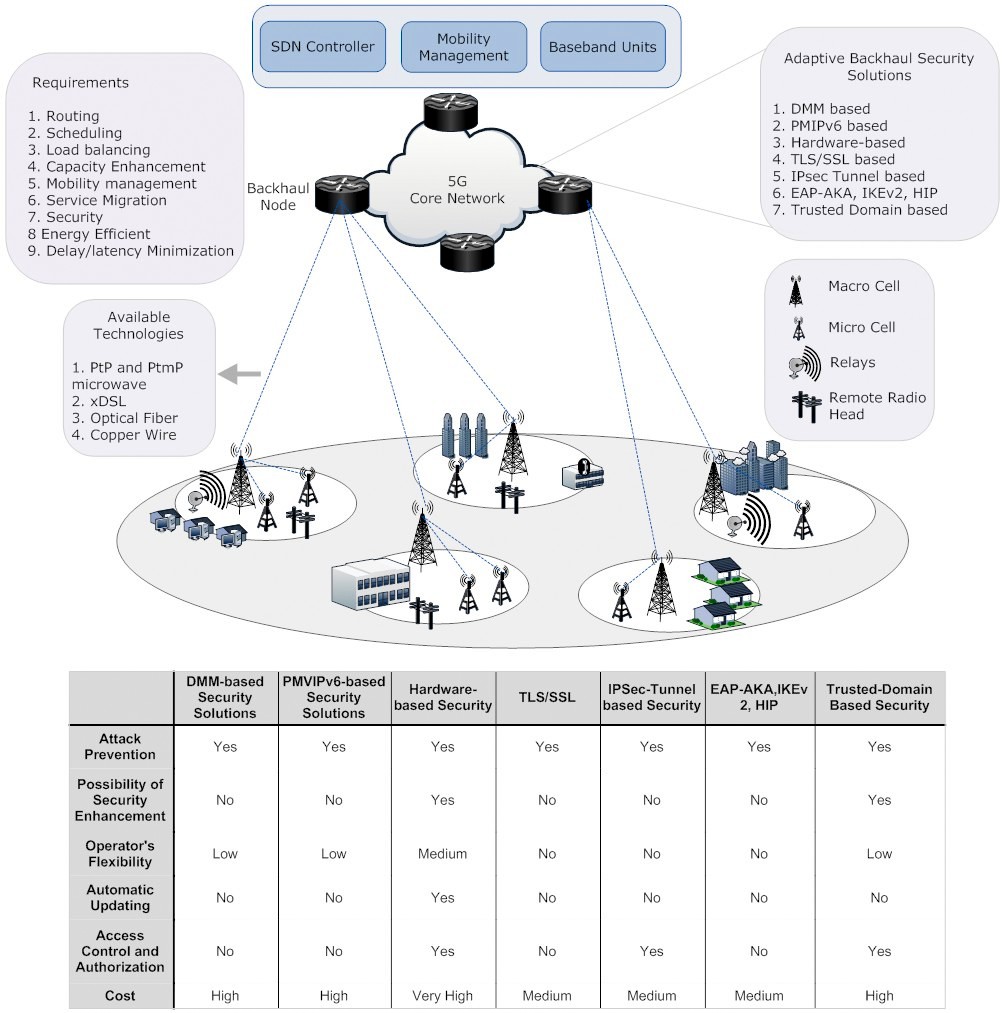}
\caption{The generalized 5G mobile backhaul network architecture and applicable security solutions.}
\label{casestudy}
\end{figure}

Fig~\ref{casestudy} represents the 5G mobile backhaul architecture and available backhaul security solutions which can be applied to meet the emerging requirements of these networks. The available security solutions are focused on the requirements of both the network and operator side. The security solutions emphasized the strong demand for lower-cost-per-bit for cellular network transport and considered various factors such as cost, bandwidth utilization, the operator’s flexibility, etc. The applicability of such solution affects the performance and but requires some additional costs for network equipment.

In addition to the above discussions, the distributed security needs a dedicated synchronization to meet a low latency and high data rates. The Distributed Mobility Management (DMM) is responsible for the mobile user handovers and connectivity, therefore, DMM-based security is also a key solution for securing mobile backhaul~\cite{sharma2018secure1}. Furthermore, PMIPv6 is also an emerging security solution and widely adopted solution by standard organizations~\cite{eiza2018efficient}~\cite{sharma2019mih}. The distributed PMIPv6 is also applicable in such requirements where the control plane and user plane needs a distributed security. These security solutions need to cope-up with traffic management solutions and improve security and resolve issues related to route optimizations.

The security requirements of basic backhaul networks include threats and attacks preventions, access control and node authorizations, traffic classifications, user privacy, dynamic coordination, and secure tunnel establishments. The available solutions attempt to place all the properties into a single package, however, its cost has affected the business model of the mobile operator. The existing solutions need to consider both network and mobile operators' factors to sustain a network without affecting its performance. The cost of hardware and software security solutions put an extra burden. Therefore, such considerations should be re-emphasized in the security requirements of 5G mobile backhauls~\cite{wu2018survey}.

\section{Research Challenges and Future Directions}
\begin{figure}[ht!]
\centering
\includegraphics[width=400px]{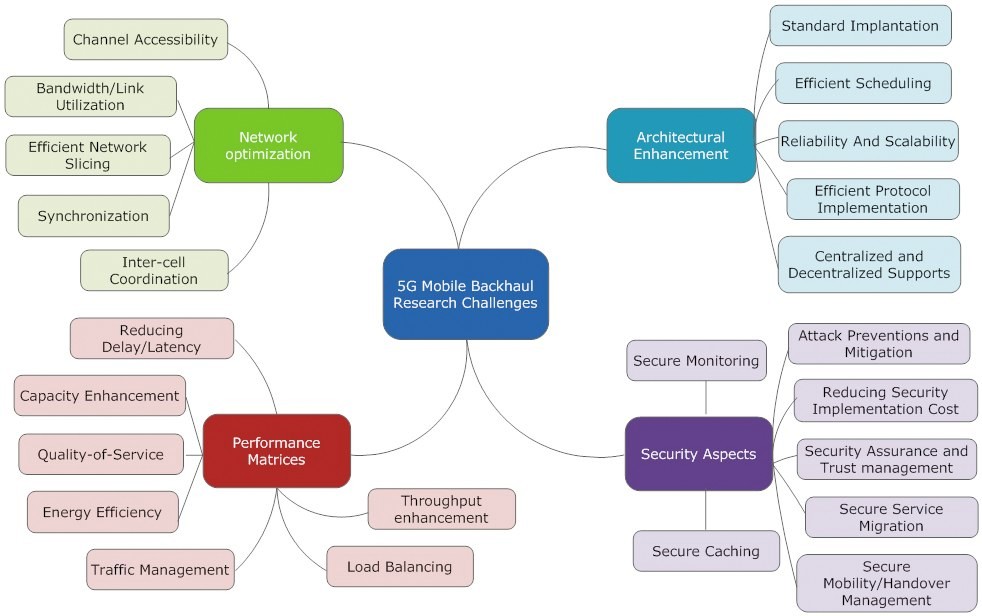}
\caption{The research challenges and future direction for 5G mobile backhaul networks.}
\label{challanges}
\end{figure}

Larger traffic is expected to flow over the 5G network with a high speed and low latency requirements. In the existing literature, security and performance are considered as the major concerns for 5G mobile backhaul networks. The existing literature presented various prominent solutions in order to enhance the performance of the 5G mobile network. Furthermore, some appropriate backhaul network architectures and protocols are required to sustain the network performance. In this section, major research challenges in the area of 5G mobile backhaul networks are discussed, as shown in Fig~\ref{challanges}.

\begin{itemize}
\item \textbf{Network Optimization}:  The network compromises various sub-modules which individually play a significant role in the trusted data transmissions. The channel interference and path loss will lead to the degradation of network efficiency. The strategically effective solutions for channel accessibility are an open issue to be solved in future researches. The link failure and node isolations affect the bandwidth utilization of the networks. The major concern which is widely considered in the articles is about the bandwidth utilization for better network formations. Furthermore, the 5G networks give the flexibility of virtual partition of the network, known as network slicing. This emerging concept helps in various application-based network allocations. Therefore, an efficient network slicing supports the backhaul networks to handle loads and efficiently manage the traffic. Additionally, the mobile operators are facing a big challenge on network synchronization. Time synchronizations are the considerable issues in the heterogeneous networks and affect the coordination of the multiple base stations of small cell. The synchronization for small cells plays a major role in the effective performance of small cells. Therefore, new synchronization techniques for 5G mobile backhaul networks need further conceptualized research by including major constraints, for instance, environmental profile, UE density per unit area, etc. The synchronization and inter-cell coordination are directly proportional to each other. The better time synchronization will lead to a better-inter-cell coordination in the small cell or ultra-dense cell. Although major enabling technologies of 5G made backhaul networks are promising, researchers still need to focus on better network formation in 5G mobile backhaul networks.

\item \textbf{Architectural Enhancement}: The architecture of a network is the main pillar for better communication and quality of service transmission. The architecture design needs to consider the factor which is ought to be centered on the performance of the network. The security standards and various key organizations are providing immense contributions to the rapid development of 5G backhaul solutions. The adoptions of standard solutions will increase the network's sustainability and enhance the communication speed. Furthermore, scheduling is one of the challenges that is often associated with the architectural phase. The Time Division Multiplexing(TDM)-based scheduling is used for the mmWave-based backhaul. An efficient scheduling helps to reduce cost and enhance the transmission rate over the network. The protocols for joint scheduling and routing for mmWave-based backhaul networks need considerable enhancement. Furthermore, the standalone solutions are restricted to some limitations. Therefore, decentralized or distributed security and service handling can be an alternative solution. The SDN/NFV-based monitoring for backhaul and traffic is expected to be an emerging trend for future research.

\item \textbf{Performance Metrics}: Various parameters are involved in the 5G communication networks which affect the performance of the network. The performance of the communication networks can be defined over the capacity, quality, and lifetime of a network, efficient routing, low latency, and high reliability. 5G networks facilitated huge traffic which needs a better load balancing capacity to sustain the networks without any failures. The network coverage can be enhanced by implementing microwave in the Point-to-Point (PTP), and Point-to-Multipoint (PMP) transmissions. The excessive data content may lead to backhaul bottleneck; therefore, the content-aware caching is needed. The global energy consumption model helps to reduce energy consumption in wired and wireless access technologies in the 5G mobile backhaul networks. The RAN and backhaul collaboration will open new doors for better results in terms of performance. The recent advancements in communication networks emphasized that the energy consumption is based on carried traffic modeling and topology selections. The radio resource management issues are widely discussed in the existing architectures. For massive traffic, new emerging technology Massive MIMO is used for a multi-user MIMO technology which enhances both capacity and implementation flexibility. The future researches should be focused on better QoS and efficient traffic management. The throughput enhancement can be achieved by efficient load balancing.

\item \textbf{Security Aspects}: Security is an important concern for the 5G networks. Trusted networks are a most demanded service in the communication networks. The privacy of a subscriber can be compromised by an attacker by exploiting vulnerabilities of the network. Various standards such as 5GPP and 3GPP emphasized the adequate security solutions to mitigate the risk factors associated with the user data. The SDN-based security solutions are widely adopted over the communication networks. However, the decentralized and distributed data creates an issue with mentoring and caching and consequently opens the opportunity for attacks. To reduce these threats, proper security monitoring and secure caching are crucial. Security of mobile backhaul should include secure link management, communication, and stable environment in the handover and mobility phases. The handover security is based on the deployed devices and wireless links between the devices~\cite{sharma2017efficient}.

\end{itemize}

The ongoing research has contributed to a significant level of advancement in the technologies and optimization in existing backhaul solutions. In any case, the existing solutions require more effective alternatives, which investigate the prerequisite of the network like performance, security, reliability and satisfy all the above-discussed requirements~\cite{wang2014cellular}. Additionally, the tradeoff between the performance and the security is needed to be managed for effective communications in the 5G mobile backhaul networks.

\section{Conclusions}
The tremendous advancements in communication technologies enhance the quality and performance of mobile networks. The mobile backhaul needs a considerable attention due to exorbitant traffic by the 5G applicants. To present details of the existing solution at one place, this article discusses various subsisting solutions for capacity and coverage of 5G-mobile backhaul networks. A generalized discussion of the existing surveys has been covered by highlighting their major contributions in the 5G mobile backhaul networks. The mobile backhaul performance factors like QoS, routing and scheduling, resource management, capacity enhancement, latency, and handovers have been discussed with SDN and mmWave based formations. Furthermore, a taxonomy of the 5G mobile backhaul framework and solutions, which includes general solutions, SDN-based solution, and the mmWave-based solution, has been discussed. The state-of-the-art comparison on the basis of various parameters of existing frameworks has been presented. It is conclusively evident that the security of mobile networks needs an attention due to an increase in threats and attacks. To further understand these implications, the possible security solutions for mobile backhaul have been discussed through various affecting factors. The trails of research challenges and future directions have been presented for the upcoming solutions which should revolve around the currents issues and overcome the subsisting constraints of 5G mobile backhaul networks.

%------------------------------------------------------------------------------
%\subsection*{Acknowledgments}
%\label{sect:acks}

\section*{Acknowledgement}
This work was supported by ‘The Cross-Ministry Giga KOREA Project’ grant funded by the Korea government (MSIT) (No. GK18N0600, Development of 20 Gbps P2MP wireless backhaul for 5G convergence service) and by the Soonchunhyang University Research Fund.

%------------------------------------------------------------------------------

%------------------------------------------------------------------------------
% Refs:
%
\label{sect:bib}
\bibliographystyle{abbrv}
\bibliography{related_works}
%\nocite{*}
%------------------------------------------------------------------------------
\section*{Author Biography}
\vspace*{1em}

\begin{biography}{Gaurav Choudhary}{a1} received the B.Tech. degree in Computer Science and Engineering from Rajasthan Technical University in 2014 and the Master Degree in Cyber Security from Sardar Patel University of Police in 2017. He is currently pursuing Ph.D. degree in the Department of Information Security Engineering, Soonchunhyang University, Asan, South Korea. His areas of research and interests are UAVs, Mobile and Internet security, IoT security, Network security, and Cryptography.
\end{biography}

\vspace*{0.5em}
%------------------------------------------------------------------------------
\begin{biography}{Jiyoon Kim}{a5} received the B.S. degree in information security engineering from Soonchunhyang University, Asan, South Korea, where he is currently pursuing the master degree in the Department of Information Security Engineering. His current research interests include mobile Internet security, IoT security, and formal security analysis.
\end{biography}

\vspace*{3em}

\vspace*{0.5em}
%------------------------------------------------------------------------------
\begin{biography}{Vishal Sharma}{a3} received the Ph.D. and B.Tech. degrees in computer science and engineering from Thapar University (2016) and Punjab Technical University (2012), respectively. He worked at Thapar University as a Lecturer from Apr’16-Oct’16. From Nov. 2016 to Sept. 2017, he was a joint post-doctoral researcher in MobiSec Lab. at Department of Information Security Engineering, Soonchunhyang University, and Soongsil University, Republic of Korea. Dr. Sharma is now a Research Assistant Professor in the Department of Information Security Engineering, Soonchunhyang University, The Republic of Korea. Dr. Sharma received three best paper awards from the IEEE International Conference on Communication, Management and Information Technology (ICCMIT), Warsaw, Poland in April 2017; from CISC-S’17 South Korea in June 2017; and from IoTaas Taiwan in September 2017. He is the member of IEEE, a professional member of ACM and past Chair for ACM Student Chapter- TU Patiala. He serves as the program committee member for the Journal of Wireless Mobile Networks, Ubiquitous Computing, and Dependable Applications (JoWUA). He was the track chair of MobiSec’16 and AIMS-FSS’16, and PC member and reviewer of MIST’16 and MIST’17, respectively. He was the TPC member of ITNACIEEE TCBD’17 and serving as TPC member of ICCMIT’18, CoCoNet’18 and ITNAC-IEEE TCBD’18. Also, he serves as a reviewer for various IEEE Transactions and other journals. His areas of research and interests are 5G networks, UAVs, estimation theory, and artificial intelligence.
\end{biography}
%------------------------------------------------------------------------------
\vspace*{0.5em}
%------------------------------------------------------------------------
%------------------------------------------------------------------------------
\end{document}